\documentclass[12pt]{iopart}
\usepackage{graphicx}
\usepackage{iopams}
\begin{document}

\setcounter{footnote}{1}

\topical{Nuclear shadowing}

\author{N\'estor Armesto}

\address{Departamento de F\'{\i}sica de Part\'{\i}culas and
IGFAE,
Universidade de Santiago de Compostela,
 15782 Santiago de Compostela, Spain}
\ead{nestor@fpaxp1.usc.es}
\begin{abstract}
The phenomenon of shadowing of nuclear structure functions at small values of
Bjorken-$x$ is analyzed. First,
multiple scattering is discussed as the underlying physical
mechanism. In this context three different but related
approaches are presented: Glauber-like rescatterings,
Gribov inelastic shadowing and ideas based on high-density Quantum
Chromodynamics.
Next, different parametrizations of nuclear
partonic distributions based on fit analysis to existing data combined
with Dokshitzer-Gribov-Lipatov-Altarelli-Parisi evolution, are reviewed.
Finally, a
comparison of the different approaches is shown,
and a few phenomenological consequences of
nuclear shadowing in high-energy nuclear collisions are presented.
\end{abstract}

\maketitle

\section{Introduction}
\label{intro}

\begin{figure}[htb]
\begin{center}
\includegraphics[width=12cm]{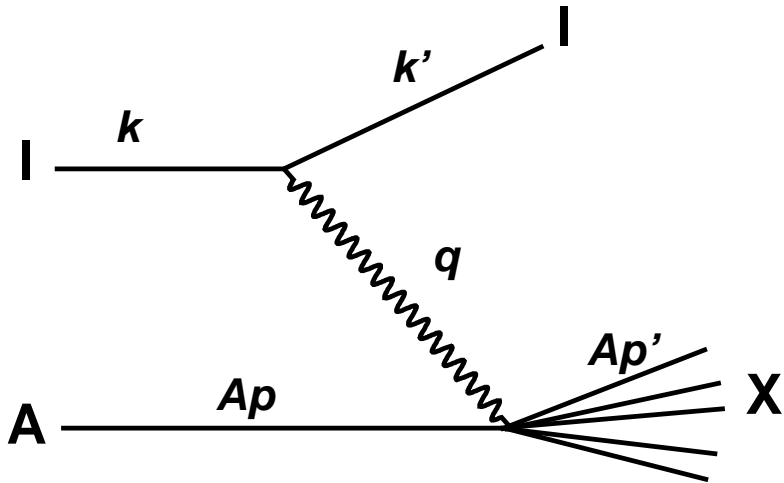}
\vskip -1cm
\caption{Diagram of leptoproduction on a nucleus
through virtual photon exchange.}
\label{fig1}
\end{center}
\end{figure}

The fact that nuclear structure functions in nuclei are different from the
superposition of those of their constituents nucleons is a well known
phenomenon since the early seventies, see references in the
reviews~\cite{Arneodo:1992wf,Geesaman:1995yd}. For example, for $F_2$
the nuclear
ratio is defined as the nuclear structure function per nucleon divided by
the nucleon structure function, 
\begin{equation}
R_{F_2}^A(x,Q^2)={F_2^A(x,Q^2) \over A \, F_2^{\rm nucleon}(x,Q^2)}\,.
\label{eq1}
\end{equation}
Here\footnote{Sometimes the ratio of nuclear ratios is used e.g.
$R(A/B)=R_{F_2}^A/R_{F_2}^B$.},
$A$ is the nuclear mass number (number of nucleons in the
nucleus).
The variables $x$ and $Q^2$ are defined as usually in leptoproduction or
deep inelastic scattering (DIS)
experiments:
in the scattering of a
lepton with four-momentum $k$ on a nucleus with four-momentum $Ap$ mediated
by photon exchange (the dominant process at $Q^2 \ll m_{{\rm Z}^0}^2,
m_{\rm W}^2$ where most
nuclear data exist),
$$ l(k)+A(Ap)\longrightarrow l(k^\prime) +X (Ap^\prime),$$
\begin{equation}
q=k-k^\prime,\ \ W^2=(q+p)^2,\ \
x={-q^2\over 2p\cdot q}={-q^2\over W^2-q^2-m_{\rm nucleon}^2},
\label{eq2}
\end{equation}
see Fig.~\ref{fig1}. The variable
$x$ has the meaning of the momentum fraction of the
nucleon in the nucleus carried by the parton with which the
photon has interacted. $Q^2=-q^2>0$
represents the squared inverse resolution of the photon as a probe
of the nuclear content. And $W^2$ is the
center-of-mass-system energy of the virtual
photon-nucleon collision (lepton masses have been neglected and $m_{\rm
nucleon}$ is the nucleon mass),
see e.g.~\cite{roberts} for full explanations.
The nucleon structure function is usually defined through measurements on
deuterium, $F_2^{\rm nucleon}=F_2^{\rm deuterium}/2$,
assuming nuclear effects in deuterium to be negligible.

\begin{figure}[htb]
\begin{center}
\includegraphics[width=12cm]{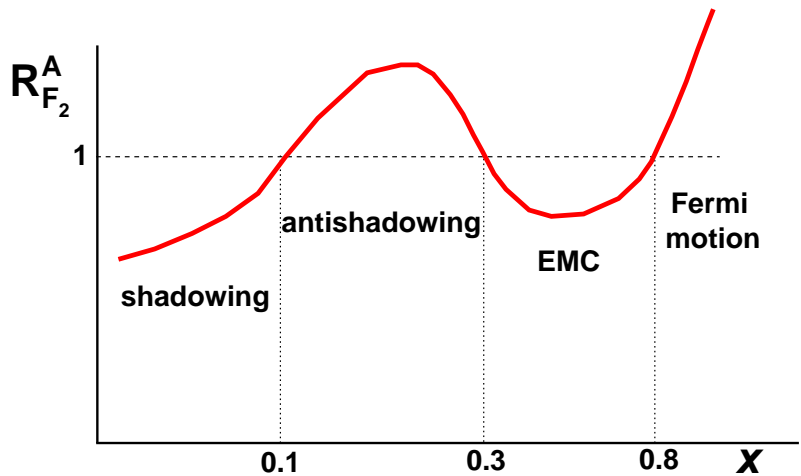}
\vskip -1cm
\caption{Schematic behaviour of $R_{F_2}^A(x,Q^2)$ as a function of $x$ for a
given
fixed $Q^2$.}
\label{fig2}
\end{center}
\end{figure}

The behaviour of $R_{F_2}^A(x,Q^2)$ as a function of $x$ for a given
fixed $Q^2$ is shown schematically
in Fig.~\ref{fig2}. It can be divided into four
regions\footnote{Note
that the deviation of the nuclear $F_2$-ratios from one in all
four regions of $x$, is sometimes referred to as the EMC effect. I use
this notation only for the depletion observed for $0.25\div 0.3 \lesssim
x\lesssim 0.8$.}:
\begin{itemize}
\item $R_{F_2}^A>1$ for $x\gtrsim 0.8$: the Fermi motion region.
\item $R_{F_2}^A<1$ for $0.25\div 0.3 \lesssim
x\lesssim 0.8$: the EMC region (EMC stands for European
Muon Collaboration).
\item $R_{F_2}^A>1$ for $0.1\lesssim
x\lesssim 0.25\div 0.3$: the antishadowing region.
\item $R_{F_2}^A<1$ for $
x\lesssim 0.1$: the shadowing region.
\end{itemize}

\begin{figure}[htb]
\begin{center}
\includegraphics[width=7.8cm]{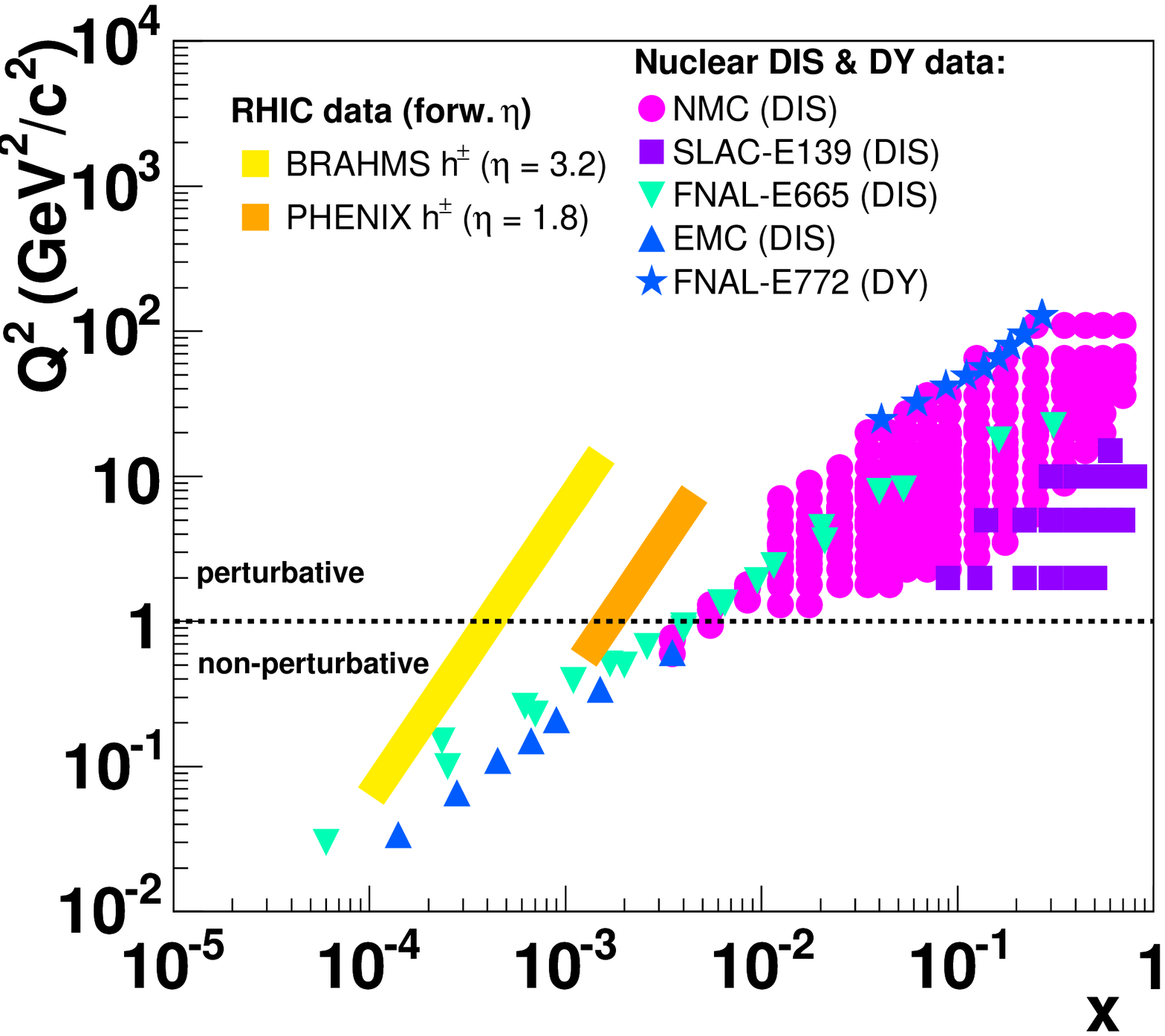}\includegraphics[width=7.8cm]{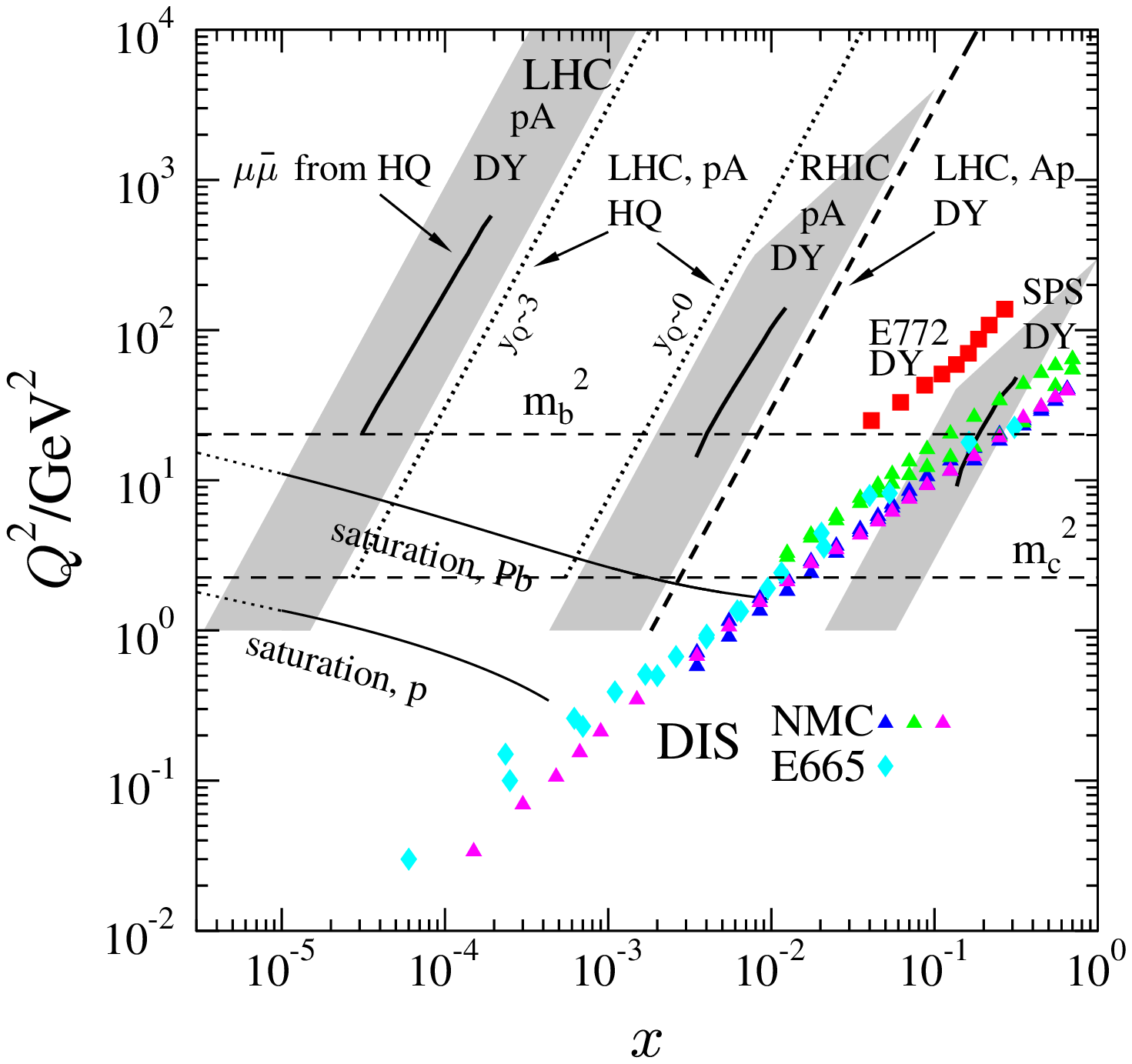}
\caption{Plot on the left: Kinematical range in the $x$-$Q^2$ plane probed in
nuclear
DIS~\cite{Amaudruz:1995tq,Arneodo:1995cs,Arneodo:1996rv,Arneodo:1996ru,Adams:1992nf,Adams:1995is,Arnold:1983mw,Arneodo:1988aa,Arneodo:1989sy}
and Drell-Yan~\cite{Alde:im} processes, and in d-Au at forward
rapidities~\cite{Arsene:2004ux,Adler:2004eh} at RHIC.
[Figure taken from~\cite{d'Enterria:2004nv}.]
Plot on the right:
The average values of $x$ and $Q^2$ of the DIS data from the New Muon
Collaboration~\cite{Amaudruz:1995tq,Arneodo:1995cs,Arneodo:1996rv,Arneodo:1996ru}
(triangles) and
E665~\cite{Adams:1992nf,Adams:1995is} (diamonds) in $l$-$A$, and of $x_2$
and $M^2$ of the Drell-Yan dilepton data~\cite{Alde:im} (squares) in
p-$A$. The heavy quark mass scales are shown by the horizontal dashed
lines. Those lines labeled saturation indicate the estimated
saturation scale in proton
and Pb.
The different bands and lines show the values of $x$ and $Q^2$ which are
or will be probed in Drell-Yan or heavy flavour production at SPS, RHIC and
LHC, for rapidities different from central ones when indicated.
[Figure taken from~\cite{Accardi:2003be}.]
See also the text in Subsection~\ref{hdqcd} and in
Section~\ref{appli}.}
\label{fig3}
\end{center}
\end{figure}

This review will be focused in the small $x$ region i.e. that of shadowing,
see~\cite{Arneodo:1992wf,Geesaman:1995yd} for discussions on the other
regions\footnote{The region of Fermi motion is
explained by the Fermi motion of the nucleons. For the EMC region there exist
several explanations: nuclear binding, pion exchange, a change in
the nucleon radius,$\dots$ The antishadowing region is usually discussed as
coming from the application of sum rules
for momentum, baryon number,$\dots$}.
The most recent experimental
data~\cite{Amaudruz:1995tq,Arneodo:1995cs,Arneodo:1996rv,Arneodo:1996ru,Adams:1992nf,Adams:1995is}
(see~\cite{Arneodo:1992wf,Geesaman:1995yd,Arnold:1983mw,Arneodo:1988aa,Arneodo:1989sy}
for previous experimental results),
confined to a limited region of
not very low $x$ and small or moderate $Q^2$ (and with a strong kinematical
correlation
between small $x$ and small $Q^2$, see Fig.~\ref{fig3}),
indicate that: i) shadowing
increases with decreasing $x$, though at the smallest available values of $x$
the behaviour is compatible with either a saturation or a mild
decrease~\cite{Adams:1992nf}; ii)
shadowing increases with the mass number of the nucleus~\cite{Arneodo:1996rv};
and iii) shadowing
decreases with increasing $Q^2$~\cite{Arneodo:1996ru}.
On the other hand, the existing
experimental data do not allow a determination of the dependence of shadowing
on the centrality of the collision.

In the region of small $x$, partonic distributions are dominated by sea quarks
and gluons. Thus isospin effects, partially corrected in practice
by the use of
deuterium as reference and of isoscalar nuclei,
are negligible and will not be discussed in the
following.
In most approaches, the origin of
the depletion of the nuclear ratios in this region
is related with
the hadronic behaviour of the virtual photon~\cite{Brodsky:1972vv}.
This resolved hadronic
component of the photon wave function at high collision energies
- equivalent to small values of $x$, see (\ref{eq2}) -
and at relatively low values of $Q^2$,
will interact several times with the different nucleons in
the nucleus i.e. will experience {\it multiple scattering}. As I will discuss
in the next Section, this results in a reduction of the corresponding cross
sections - shadowing, related to the structure functions through
\begin{equation}
F^A_2(x,Q^2)={Q^2(1-x)\over 4 \pi^2 \alpha_{\rm EM}}\,\sigma_{\gamma^*-A}
\,,
\label{eq3}
\end{equation}
with $\alpha_{\rm EM}$ the fine structure constant. Thus, the phenomenon of
multiple scattering is sometimes referred to as {\it shadowing corrections}.

The importance of the phenomenon of nuclear shadowing is twofold:
First,
on the theoretical side it offers an experimentally accessible testing ground
for our understanding of Quantum Chromodynamics (QCD)
in the high-energy regime~\cite{cargese}.
Multiple scattering is unavoidable in a quantum field theory as a consequence
of such a basic requirement of the theory as unitarity.
The nuclear size gives the possibility to control the amount of
multiple scattering at given values of momentum fraction
$x$ and scale $Q$. Besides, by
varying the scale and the energy of the collision the interplay between the
soft non-perturbative and the hard perturbative regimes can be addressed.
Second, experimental studies on high-energy nuclear collisions like those
at the Relativistic Heavy Ion Collider
(RHIC)~\cite{Adcox:2004mh,Back:2004je,Arsene:2004fa,Adams:2005dq} at the
Brookhaven National Laboratory (BNL) are presently carried out.
New possibilities like the Large Hadron Collider (LHC)~\cite{Accardi:2003be} at
CERN or the Electron-Ion Collider (EIC)~\cite{Deshpande:2005wd}
under consideration, will become available in the
future. They
test the behaviour of parton densities inside nuclei at larger
energies/smaller $x$ than those presently available in fixed target studies
like those at the Super Proton Synchrotron (SPS) at CERN. These new experimental
data offer
the possibility
to further constrain our knowledge on the behaviour of nuclear
cross sections and
structure functions, both for observables characterized by a large
scale for which standard perturbation theory can be applied, and for those
with intermediate and small scales where new methods have been developed.

The plan of the review is the following: In Section~\ref{multi} models based
on multiple scattering will be reviewed. In Section~\ref{dglap} those models
which do not try to address the origin of nuclear shadowing but rather to
study its evolution through the Dokshitzer-Gribov-Lipatov-Altarelli-Parisi
(DGLAP) equations~\cite{Dokshitzer:1977sg,Gribov:1972ri,Altarelli:1977zs},
will be discussed. A comparison of the different models
will be shown in Section~\ref{compa}.
Next,
some consequences on high-energy nuclear
collisions will be presented in Section~\ref{appli}. Finally, in
Section~\ref{conclu} some conclusions will be drawn.
I have mainly
focused on the most recent approaches roughly starting from
the early nineties,
please have a look at~\cite{Arneodo:1992wf}
for an
extensive list of references on earlier theoretical and experimental work.

\section{Models based on multiple scattering}
\label{multi}

As commented in the Introduction, the usual explanation for the origin of
shadowing is multiple
scattering~\cite{Stodolsky:1966am,Schildknecht:1973gi,Nikolaev:1975vy,Shaw:1989mn,Kulagin:1994fz,Kwiecinski:1988ys,Frankfurt:1988zg,Brodsky:1989qz,Nikolaev:1990ja,Barone:1992ej,Kopeliovich:1995yr,Armesto:1996id,Huang:1997ii,Armesto:2002ny,Kopeliovich:1999am,Kopeliovich:2000ra,Nemchik:2003wx,Capella:1997yv,Frankfurt:2002kd,Frankfurt:2003zd,Armesto:2003fi,Gribov:1984tu,Mueller:1985wy,Qiu:1986wh,Berger:1988ea,Close:1988xw,Qiu:2003vd,Castorina:2004ye}.
While the basis of the explanation is common, phenomenological details of its
application vary from model to model. For
example, the hadronic component of the virtual photon may be given a partonic
structure like in the dipole model~\cite{Nikolaev:1990ja}, see the next
Subsection, or
modeled~\cite{Stodolsky:1966am,Schildknecht:1973gi,Shaw:1989mn,Kulagin:1994fz,Kwiecinski:1988ys}
as a superposition of hadronic states with the photon
quantum numbers - vector meson dominance, or some combination of both
approaches e.g.~\cite{Frankfurt:1988zg}.
Besides, what is seen as multiple scattering in the rest frame of the nucleus
corresponds to recombination in the infinite momentum
frame~\cite{Gribov:1984tu,Mueller:1985wy}.
Just to mention a few differences between the results of the models,
the behaviour of
models~\cite{Nikolaev:1990ja,Barone:1992ej,Kopeliovich:1995yr} is dominated by
hadronic configurations of large size, so the results turns out to be basically
$Q^2$-independent. On the other hand, models which consider an expansion in
power-suppressed corrections in $1/Q$, either a fixed number of
terms~\cite{Mueller:1985wy,Qiu:1986wh,Berger:1988ea} or some
re-summation~\cite{Qiu:2003vd}, show a clear $Q^2$ dependence\footnote{The
experimental evidence of a $Q^2$ dependence of the
nuclear $F_2$-ratios comes from~\cite{Arneodo:1996ru},
thus being subsequent to many
models e.g. the analysis in~\cite{Smirnov:1995vw};
previous data did not show any clear $Q^2$-dependence.}.
Some models rely on an eikonal
approximation~\cite{Armesto:1996id,Armesto:2002ny,Qiu:2003vd}, see below, and
are unable to reproduce the return of $F_2$
nuclear ratios to 1 at $x\sim 0.1$,
while
others~\cite{Kopeliovich:1999am,Kopeliovich:2000ra,Nemchik:2003wx,Capella:1997yv,Frankfurt:2002kd,Frankfurt:2003zd,Armesto:2003fi}
include effects of finite coherence length, see below, and are able 
to reproduce such a behaviour.

In this Section I will start by working out
a little exercise which shows how multiple
scattering leads to shadowing. This exercise should also clarify the origin of
coherence effects. Then, in the Subsections models based on Glauber-like
rescatterings, on Gribov inelastic shadowing, and finally the ideas based on
high-energy QCD~\cite{cargese}, will be reviewed.

\begin{figure}[htb]
\begin{center}
\includegraphics[width=15cm,height=9cm]{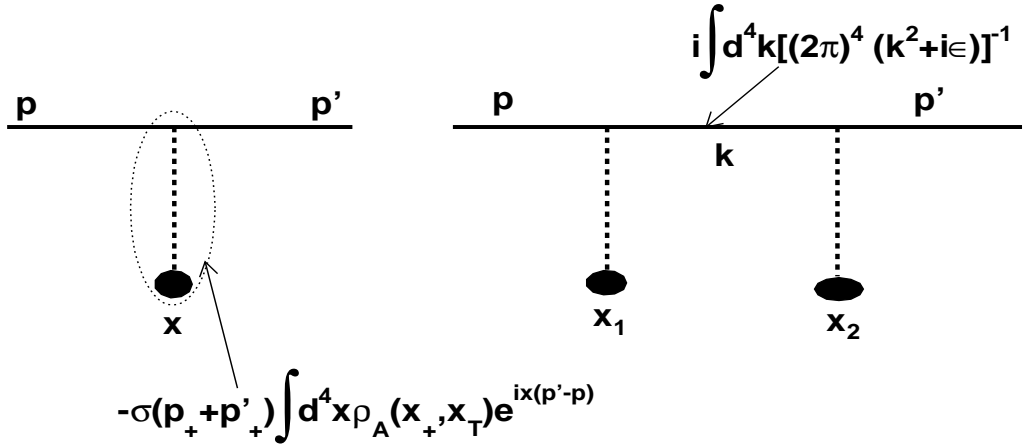}
\vskip -3cm
\caption{One- (left) and two- (right) scattering diagrams, with the
corresponding Feynman
rules written on them.}
\label{fig4}
\end{center}
\end{figure}

For the exercise I consider the contribution coming from one and two
scatterings, to the high-energy cross
section of a massless
scalar particle on a nucleus with mass number $A$.
The scattering centers plus the interaction vertex are represented by
the projectile-nucleon forward scattering amplitude times the nuclear density
(see Fig.~\ref{fig4}).
This example follows the
spirit of the Glauber-Gribov
theory~\cite{glauber,Gribov:1968jf,Gribov:1968gs};
technical details can be found in Section 3.1 and Appendix A
in~\cite{Hebecker:1999ej} for the case of scalar
QCD at high energy.
I will use light-cone coordinates $a_\pm =a_0\pm a_z$,
$a=(a_0,a_T,a_z)=(a_+,a_-,a_T)$ with $a_T=(a_x,a_y)$ the two-dimensional
transverse vector, assume dominance of
the $+$-components for the projectile, and define $q=p^\prime
-p$. I will employ the optical theorem for purely imaginary amplitudes, $i {t}
(q=0)=i t_{\rm forw}
=-\sigma$ for projectile-nucleon and $i {\cal T}_{n}(q=0)=-\sigma_A^n$ for the
$n$-scattering contribution for projectile-nucleus collisions. Then
the amplitude with one scattering (Fig.~\ref{fig4} left) reads:
\begin{eqnarray}
c(p_+,p^\prime_+) i{\cal T}_1(q)&=&i{t}
_{\rm forw}\ A(p_++p^\prime_+) \int d^4x \,
\rho_A(x_+,x_T) e^{i x\cdot (p^\prime-p)} \nonumber \\
&=& i{t}
_{\rm forw}\ c(p_+,p^\prime_+) A
\int d^2x_T\, T_A(x_T) e^{-i x_T\cdot
(p_T^\prime -p_T)}.
\label{eq4}
\end{eqnarray}
$\rho_A(x_+,x_T)$ is the nuclear density normalized to 1,
\begin{equation}
T_A(x_T)=\int_{-\infty}^{+\infty} dx_+ \,\rho_A(x_+,x_T)
\label{eq5}
\end{equation}
the nuclear profile, $|x_T|=b$ the impact parameter
and $c(p_+,p^\prime_+)=(2 \pi) 2p_+ \delta(p^\prime_+-p_+)$
a normalization factor. For the forward scattering case $q=0$, (\ref{eq4})
gives
\begin{equation}
\sigma_A^1=A \,\sigma
\label{eq6}
\end{equation}
as expected. The two-scattering contribution (Fig.~\ref{fig4} right) reads
\begin{eqnarray}
&&c(p_+,p^\prime_+) i{\cal T}_2(q)=i A(A-1) (i{t}
_{\rm forw})^2 \int {d^4 k\over (2\pi)^4} d^4 x_1 d^4 x_2\,
e^{i x_1\cdot (k-p)}
\nonumber \\
&&\times e^{i x_2\cdot
(p^\prime-k)}{(p_++k_+)
(k_++p^\prime_+) \over k^2+i\epsilon}
\rho_A(x_{1+},x_{1T}) \rho_A(x_{2+},x_{2T}) \nonumber \\
&&= c(p_+,p^\prime_+) A(A-1) (i{t}
_{\rm forw})^2 \label{eq7}\\
&&\times\int {d^2 k_T\over (2\pi)^2} dx_{1+} dx_{2+} d^2x_{1T} d^2x_{2T}\,
e^{-i k_T^2(x_{2+}-x_{1+})/(2 p_+)} \nonumber \\
&&\times e^{-i [x_{1T}\cdot (k_T-p_T)+ x_{2T}\cdot (p^\prime_T-k_T)]}
\rho_A(x_{1+},x_{1T}) \rho_A(x_{2+},x_{2T}) \theta(x_{2+}-x_{1+}),
\nonumber
\end{eqnarray}
with $\theta(x)$ the step function. The second equality follows from doing the
$dk_-$ integral closing the integration contour with an infinite semicircle in
the lower half-plane and simplifying the remaining $\delta$-functions.

Coherence effects are contained in the
factor $\exp{[-i k_T^2(x_{2+}-x_{1+})/(2 p_+)]}$ which can be written as
$\exp{[-i(x_{2+}-x_{1+})/l_c]}$, with $l_c=2p_+/k_T^2$ the coherence length.
In the low energy limit $p_+\to 0$ this factor leads to $i{\cal T}_2(q)\to 0$
(the same happens for contributions with more than two scatterings). Then all
rescattering corrections vanish, so the total cross section results equal to
the superposition of single scatterings (\ref{eq6}) - this limit is thus
called the incoherent limit. This would imply a
nuclear ratio equal to 1, so shadowing vanishes for low energies i.e. large
values of $x$. On the other hand, in the large energy, totally coherent
limit $p_+\to \infty$,
$\exp{[-i(x_{2+}-x_{1+})/l_c]}\to 1$ and (\ref{eq7}) gives
\begin{equation}
i{\cal T}_2(q)={A (A-1) \over 2} (i{t}
_{\rm forw})^2 \int d^2x_T\, e^{-i x_T\cdot (p_T^\prime-p_T)} T_A^2(x_T),
\label{eq8}
\end{equation}
which in the forward case gives
\begin{equation}
\sigma_A^2=-{A (A-1) \over 2} \int d^2x_T\,[T_A(x_T)\sigma]^2.
\label{eq9}
\end{equation}
This correction turns out to be negative so the cross section is smaller than
the superposition of independent collisions of the projectile with every
nucleon in the nucleus, and the nuclear ratio results
lower than one. In this way,
multiple scattering offers an explanation for
shadowing\footnote{Multiple scattering plays a
key role
in many physical processes.
For example, coherence effects in multiple scattering are widely
discussed in the context of medium-induced radiation both in Quantum
Electrodynamics and in QCD (see e.g.~\cite{Klein:1998du}
and~\cite{Baier:2000mf}
respectively, and references therein), or in heavy flavour production on
nuclear targets~\cite{Braun:1997qw,Kopeliovich:2001ee}.}.
Furthermore, from this example it becomes evident that shadowing increases
with increasing $\sigma$ and mass number $A$ (and, if the integration over
$x_T$ is not performed, increasing centrality
equivalent to decreasing $x_T$).
The cross section $\sigma$
increases with increasing energy (equivalent to
decreasing $x$) or decreasing $Q^2$
(equivalent to increasing size of the hadronic component of the virtual photon,
see the next Subsection). Most of these
features, as indicated in the Introduction, are
seen in the experimental data.
The differences between the models mentioned at the beginning of
this
Section come both from the modeling of $\sigma$ and to the way in which
multiple
scattering is considered. Both aspects are the subject of the following
Subsections.

In the coherent limit the hadronic fluctuation of the virtual photon interacts
with the target as a whole. The kinematical region where this happens can be
discussed using
a simple argument: In the laboratory frame where the nuclear target
is at rest, the lifetime of the hadronic fluctuation of a $\gamma^*$ with
squared virtuality $Q^2$ can be estimated using the uncertainty principle and
the Lorentz dilation in this frame,
\begin{equation}
\tau \sim {1 \over Q} \times {E_{\rm lab}\over Q} \simeq {W^2 \over 2m_{\rm
nucleon} Q^2} \simeq {1\over 2 m_{\rm
nucleon}x}\, ,
\label{eq10}
\end{equation}
where I have used (\ref{eq2}) for $2 m_{\rm
nucleon} E_{lab} \simeq W^2\gg Q^2$.
This lifetime increases with decreasing $x$ or increasing
energy. For the hadronic fluctuation to interact with the nuclear
target as a whole, the lifetime has to be greater than the nuclear radius,
$\tau > R_A$. This implies $x\lesssim 1/(2 m_{\rm
nucleon} R_A)$. Using typical values $R_A \sim A^{1/3}$ fm, we get $x\lesssim
0.1 A^{-1/3}$ which roughly coincides with the experimentally measured
values of $x$ for which the
transition from antishadowing
to shadowing takes place.

\subsection{Glauber-like rescatterings}
\label{glauber}

Some models try to address the origin of nuclear shadowing through the
Glauber-Gribov formalism in the totally coherent
limit~\cite{Armesto:1996id,Armesto:2002ny,Braun:1997qw}.
A proper treatment of
coherence effects requires the consideration of the mass spectrum of
intermediate fluctuations and will be discussed in detail in the next
Subsection, though some models~\cite{Huang:1997ii} include coherence in an
effective way.

The Glauber-Gribov theory~\cite{glauber,Gribov:1968jf,Gribov:1968gs} considers
the
multiple scattering of the hadronic component of the virtual photon with a
nucleus made of nucleons whose binding energy is neglected.
This hadronic component keeps a fixed size
during the scattering process - the eikonal approximation,
and is usually limited to its lowest-order Fock state, a $q\bar q$ pair - the
so-called dipole
model~\cite{Nikolaev:1990ja,Mueller:1993rr,Mueller:1994jq}. Then
the total dipole-nucleus cross section reads
\begin{equation}
\sigma_{{\rm dipole}-A}(x,r)=\int d^2b
\ \ 2\left[1-\exp{\left(-\frac{1}{2}AT_A(b)\sigma_{\rm dipole-nucleon}(x,r)
\right)}\right],
\label{eq11}
\end{equation}
with $b$ the impact parameter of the center of the dipole
relative to the center of the nucleus and $r$ the size of the dipole.
This cross section is then related to
the nuclear $F_2$ through (\ref{eq3}) and
\begin{equation}
\sigma_{\gamma^*-A}
(x,Q^2)=\int d^2r\,\rho(r,Q^2)\,\sigma_{{\rm dipole}-A}(x,r),
\label{eq12}
\end{equation}
where $\rho(r,Q^2)$ are the distributions of colour dipoles of size $r$
created by splitting of the incident photon into a $q\bar q$
pair~\cite{Nikolaev:1990ja,Mueller:1993rr,Mueller:1994jq}. These distributions
provide a definite relation between increasing $Q$ and decreasing $r$ (or vice
versa).

\begin{figure}[htb]
\begin{center}
\includegraphics[width=7.8cm]{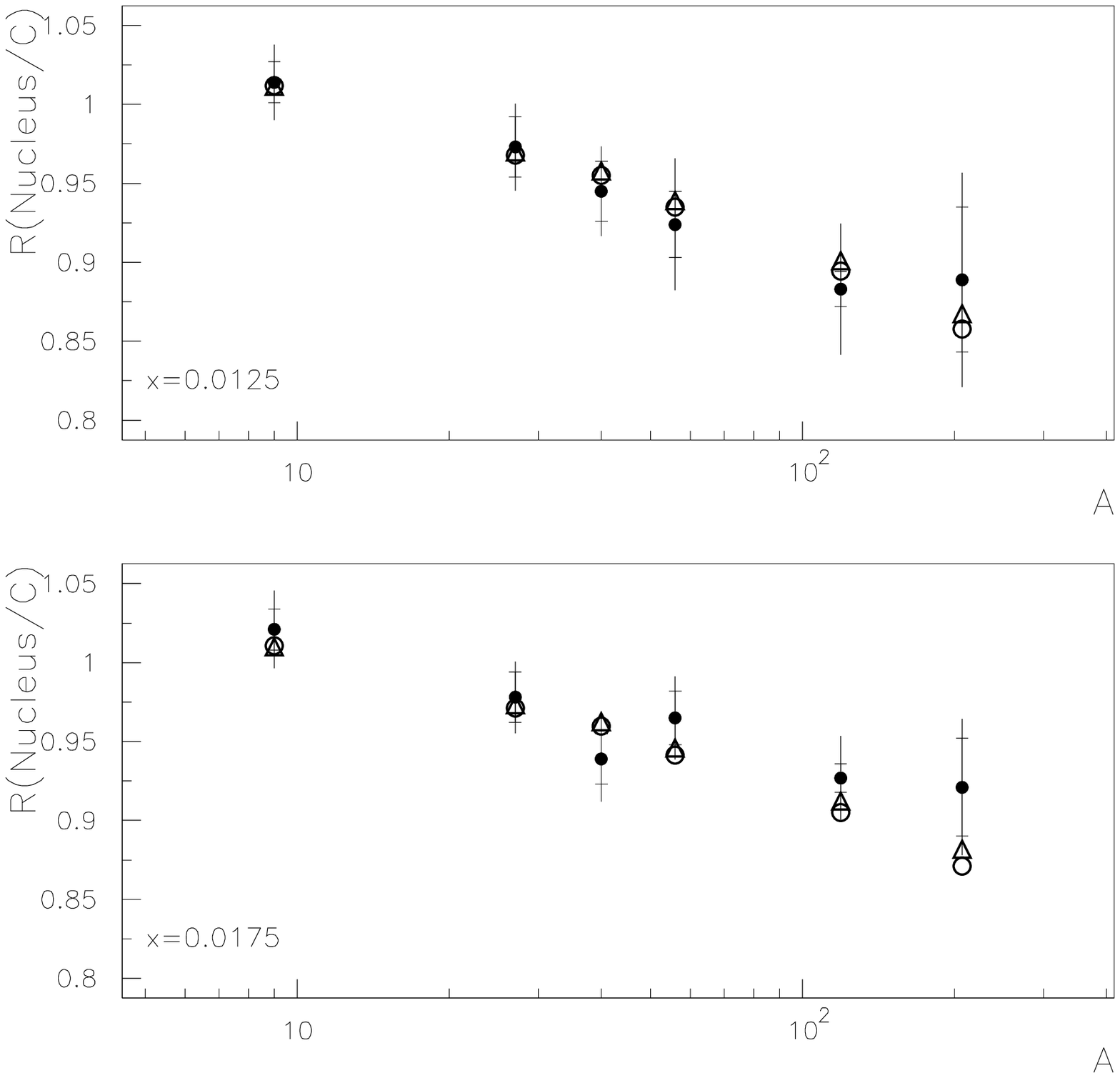}\includegraphics[width=7.8cm]{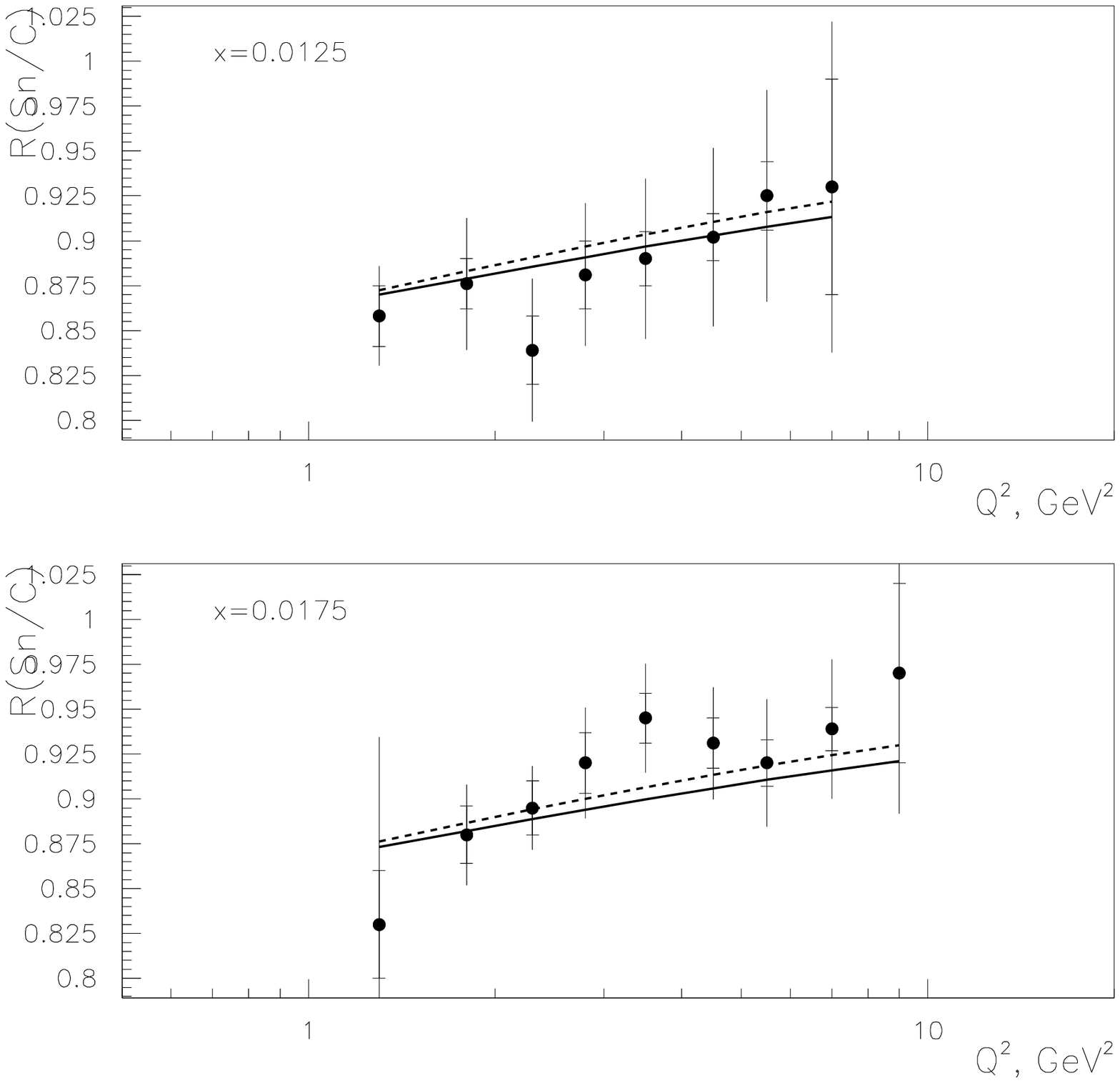}
\caption{Plots on the
left: Nuclear size dependence of the $F_2$-ratios in the Glauber model
in~\cite{Armesto:2002ny} for two fixed values of
$x$, compared with experimental data~\cite{Arneodo:1996rv} (filled points).
Plots on the right:
$Q^2$-dependence of the nuclear $F_2$-ratios for two fixed values of
$x$, compared with experimental
data~\cite{Arneodo:1996ru} (filled points). Open triangles and
circles in the plots on the left, and solid and dashed lines in the plots on
the right, correspond to different
models for
the dipole-nucleon cross section. In the experimental points, inner
error
bars correspond to statistical errors, and the outer ones to statistical plus
systematic errors added in quadrature.
[Figures taken from~\cite{Armesto:2002ny}.]}
\label{fig5}
\end{center}
\end{figure}

The differences between realizations of this approach come mainly
from the model used for the dipole-nucleon cross section (equivalent to the
dipole-proton cross section at the small values of $x$ where this approach is
applicable). For example,
in~\cite{Armesto:2002ny} a parametrization
based on a saturation model~\cite{Golec-Biernat:1998js} is used, while
in~\cite{Armesto:1996id} a form based on the Balitsky-Fadin-Kuraev-Lipatov
(BFKL)
pomeron~\cite{Kuraev:1977fs,Balitsky:1978ic}
is employed, see Subsection~\ref{hdqcd}.
On the other hand, in~\cite{Huang:1997ii} forms based on the
double-leading-log (DLL) approximation to the DGLAP evolution equations
are taken,
see~\cite{Frankfurt:1996ri} for a discussion of the relation of scales
in the dipole model and
DGLAP.
All of them give a reasonable description of the data on nuclear
shadowing for $x\lesssim 0.01$, see e.g. Fig.~\ref{fig5},
although the effective introduction of coherence
effects in~\cite{Huang:1997ii} allows to describe the whole shadowing region.
The $Q^2$-dependence of nuclear ratios~\cite{Arneodo:1996ru} is also
reproduced~\cite{Armesto:2002ny}.
It must be stressed that once the dipole-nucleon cross section is fixed, the
extension to the nuclear case is essentially parameter-free, making the
agreement with the nuclear experimental data more remarkable.

Apart from their intrinsic interest (see e.g. an application to exclusive
vector meson photoproduction in~\cite{Goncalves:2004bp}),
these models have also been used as
initial conditions at not very small $x\sim 0.01$, for evolution towards
smaller values of $x$ in the framework of high-density QCD, see
Subsection~\ref{hdqcd}.
Also there I will discuss the issues of the saturation scale
which can be extracted in this framework.

\subsection{Gribov inelastic shadowing}
\label{gribov}

\begin{figure}[htb]
\begin{center}
\includegraphics[width=13cm]{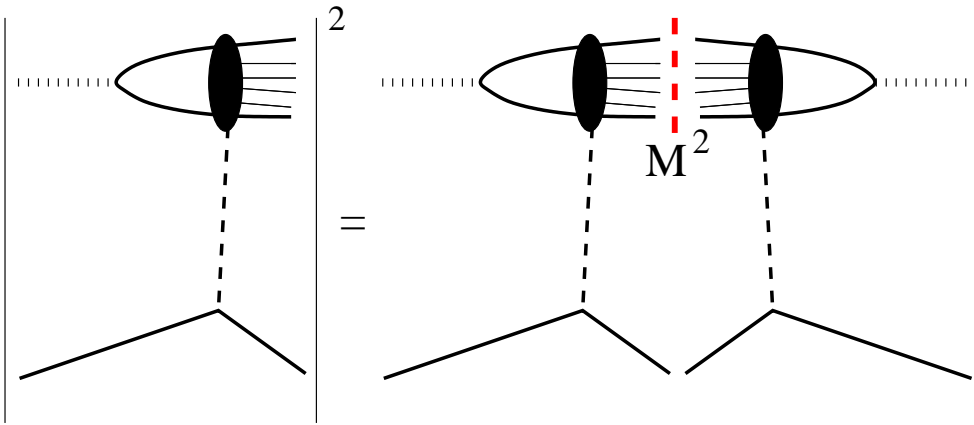}
\caption{Diagrams relating diffraction with the two-scattering contribution to
the total cross section.}
\label{fig6}
\end{center}
\end{figure}

In the classical Glauber model~\cite{glauber} subsequent interactions of the
projectile with nucleons in the nucleus occur,
and the intermediate states of the
projectile are the same as the initial one i.e. {\it elastic}. In the
relativistic Gribov theory~\cite{Gribov:1968jf,Gribov:1968gs},
subsequent interactions are suppressed at high
energies and the collision proceeds through
simultaneous interactions of the projectile with the nucleons in the nucleus.
The intermediate states are no longer the same as the initial state and are
called {\it inelastic}.
The use of Reggeon calculus~\cite{Gribov:1968fc} and the
Abramovsky-Gribov-Kancheli (AGK) cutting rules~\cite{Abramovsky:1973fm}
(see an updated discussion in~\cite{Bartels:2005wa}) allow to write a relation
between the cross section for diffractive dissociation of the projectile and
the two-scattering contribution to
the projectile-target cross section, see Fig.~\ref{fig6}. 

\begin{figure}[htb]
\begin{center}
\includegraphics[width=13cm]{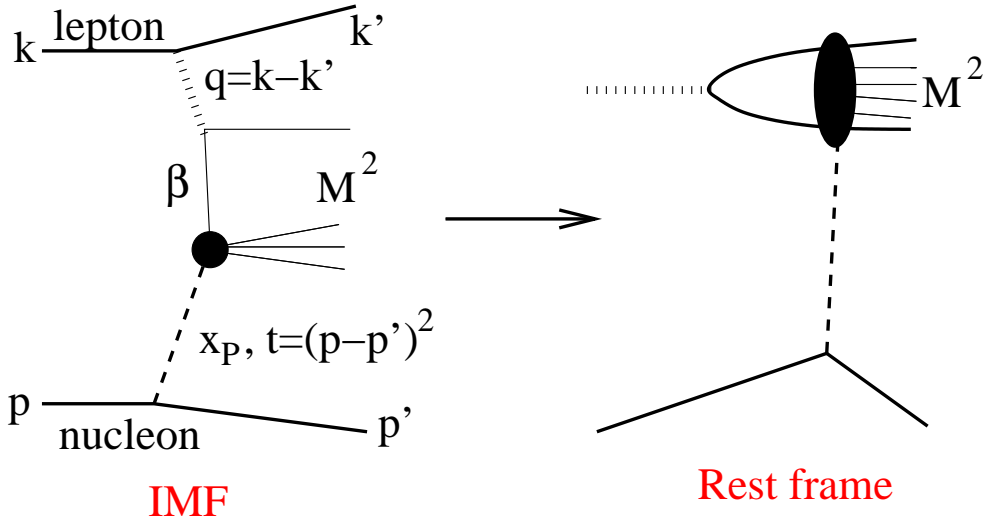}
\caption{Diffractive dissociation in the infinite momentum frame (IMF),
with the corresponding kinematical variables,
and in
the rest frame of the hadron.}
\label{fig7}
\end{center}
\end{figure}

The
first correction to the non-additivity of cross sections comes from
the second-order rescattering $\sigma_A^{2}$.
In Fig.~\ref{fig7} diffractive DIS is shown both in the infinite momentum frame
and in the rest frame of the nucleon. From Fig.~\ref{fig6}
it becomes clear that the
square of such contribution is equivalent to a double exchange with a cut
between the exchanged amplitudes, a so-called diffractive cut.
To compute the first
contribution to nuclear shadowing $\sigma_A^{2}$ which comes from these
two exchanges, one needs
its total contribution to the $\gamma^*$-nucleon cross section.
This contribution arises from cutting the two-exchange amplitude in
all possible ways: between the amplitudes and the amplitudes themselves in
all possible manners. For purely imaginary amplitudes,
it can be shown \cite{Gribov:1968fc,Abramovsky:1973fm} that this
total contribution
is identical to minus the contribution from the diffractive cut. Thus
diffractive DIS is directly related to the first contribution
to nuclear shadowing.
The final expression reads
\begin{equation}
\sigma_A^{2}=-4\pi A(A-1)\int d^2b\ T_A^2(b)
\int _{M^2_{min}}^{M^2_{max}}dM^2 \left.
\frac{d\sigma^{\mathcal{D}}_{\gamma^*-{\rm p}}}{dM^2dt}\right\vert_{t=0}
F_A^2(t_{min}),
\label{eq13}
\end{equation}
with $M^2$ the mass of the diffractively produced system, $M^2_{min}\simeq
4m_\pi^2$ and $M^2_{max}= Q^2\left(x_{Pmax}/x-1\right)$
with $x_{Pmax}\sim 0.1$, see~\cite{Armesto:2003fi}.
The usual variables for diffractive DIS: $Q^2$, $x$, $M^2$ and $t$,
or
$x_P=x/\beta$, $\beta=Q^2/(Q^2+M^2)$, are shown in Fig.~\ref{fig7}.
$d\sigma^{\mathcal{D}}_{\gamma^*-{\rm p}}/dM^2dt$ is the
differential cross section for diffractive dissociation of the virtual photon.
Coherence effects
are taken into account through
\begin{equation}
F_A(t_{min})=\int d^2b\ J_0(b\sqrt{-t_{min}})T_A(b),
\label{eq14}
\end{equation}
with $t_{min}=-m_{\rm nucleon}^2x_P^2$.
This function is equal to 1 at $x\to 0$ and
decreases
with increasing $x$ due to the loss of coherence for $x\gtrsim
(2m_{\rm nucleon}R_A)^{-1}$, see (\ref{eq10}).
Details can be found in the corresponding
references~\cite{Kopeliovich:1999am,Kopeliovich:2000ra,Nemchik:2003wx,Capella:1997yv,Frankfurt:2002kd,Frankfurt:2003zd,Armesto:2003fi}.

The differences between available realizations come from the consideration of
real parts in the pomeron amplitude~\cite{Frankfurt:2002kd,Frankfurt:2003zd},
or from the model or parametrization used for the diffractive cross section:
phenomenological models and
parametrizations~\cite{Capella:1997yv,Frankfurt:2002kd,Frankfurt:2003zd,Armesto:2003fi}
which reproduce the existing experimental data\footnote{The use of such
parametrizations can be justified at large enough $Q^2$ by the factorization
theorem for diffractive hard scattering~\cite{Collins:1997sr}.},
or a model which considers the
evolution of a diffractive $q\bar q$ state
in~\cite{Kopeliovich:1999am,Kopeliovich:2000ra,Nemchik:2003wx} (see the
method in~\cite{Zakharov:1998sv}).
Differences also come from the
extension of the models to include higher order rescatterings. Such extensions
are model-dependent. Some models consider
that all intermediate states in the
subsequent rescatterings have the same
structure~\cite{Capella:1997yv,Frankfurt:2002kd,Frankfurt:2003zd,Armesto:2003fi},
in the form of a Schwimmer~\cite{Schwimmer:1975bv}
\begin{equation}
\label{eq15}
\sigma^{Sch}_{\gamma^*-A}=\sigma_{\gamma^*-{\rm nucleon}}\int d^2b
\,\frac{AT_A(b)}{1+(A-1)f(x,Q^2)T_A(b)}
\end{equation}
or an eikonal unitarized cross section
\begin{equation}
\label{eq16}
\sigma^{eik}_{\gamma^*-A}=\sigma_{\gamma^*-{\rm nucleon}} \int d^2b
\,\frac{A\left\{1-\exp{\left[-2(A-1)T_A(b)f(x,Q^2)\right]
}\right\}}{2(A-1)f(x,Q^2)}
\end{equation}
(with $f(x,Q^2)$ defined to get
consistency with (\ref{eq13}) when these equations are
expanded to second order). Other models take into account
the possibility of different intermediate
states~\cite{Kopeliovich:1999am,Kopeliovich:2000ra,Nemchik:2003wx}.

\begin{figure}[htb]
\begin{center}
\includegraphics[width=14cm]{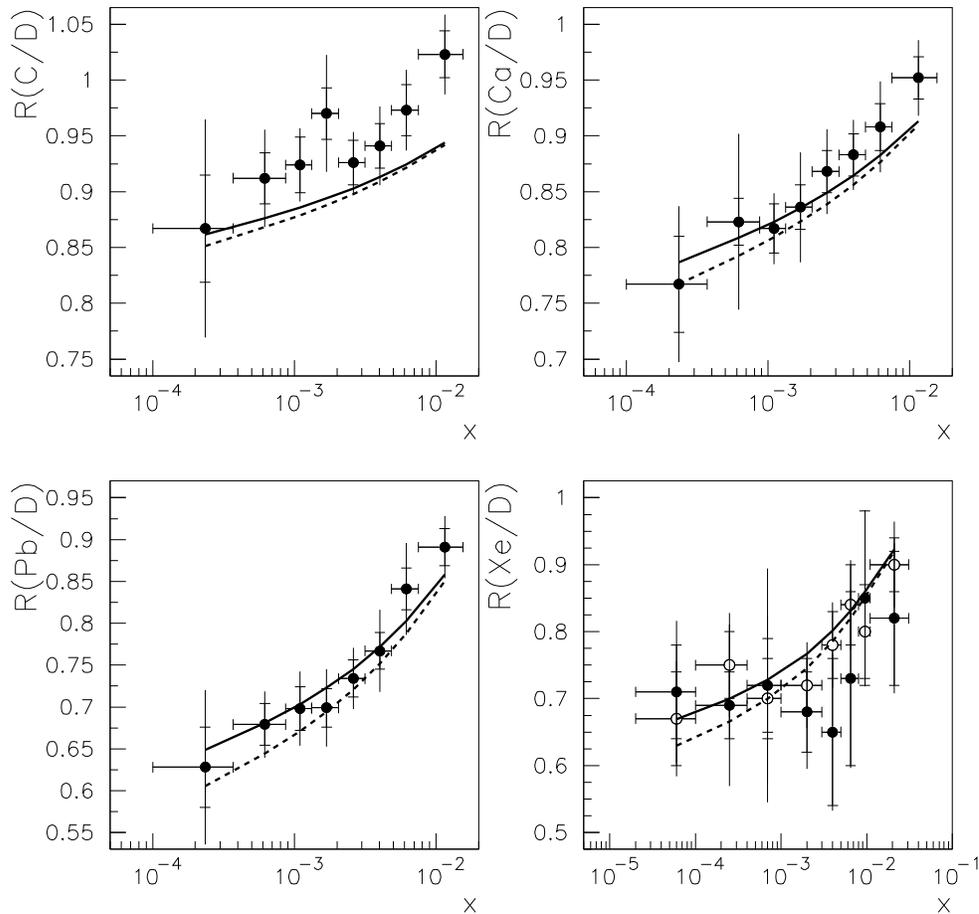}
\caption{$x$-dependence of the nuclear $F_2$-ratios in the model
in~\cite{Armesto:2003fi} for different nuclei,
compared with experimental data~\cite{Adams:1992nf,Adams:1995is} (circles).
Solid lines correspond to Schwimmer (\ref{eq15}) and
dashed lines to eikonal
(\ref{eq16}) unitarization. Error bars in the experimental points follow the
same convention as in Fig.~\ref{fig5}.
In the ratios
Xe/D, filled circles correspond to the
analysis with hadron requirement and open circles to that with
electromagnetic cuts, see the experimental paper~\cite{Adams:1992nf}
for more details. Both the experimental results and the theoretical ones,
joined by lines, correspond to different average $Q^2$ for every different
value of $x$.
[Figure taken from~\cite{Armesto:2003fi}.]}
\label{fig8}
\end{center}
\end{figure}

Several comments are in order. First, once the model or parametrization for the
diffractive cross section is provided, the extension to the nuclear case is
parameter-free - except for the modeling of higher order rescatterings
(\ref{eq15}) and (\ref{eq16}). These models, together with those in
Subsection~\ref{glauber}, provide an impact parameter dependence of nuclear
shadowing. In this way, this approach and the one
presented in the previous Subsection offer a link between the nucleon and
nuclear cases. Second, these models can be used as initial conditions for
DGLAP evolution as done
in~\cite{Frankfurt:2002kd,Frankfurt:2003zd}, see the next
Section. Third, both models for the diffractive cross section and their
extension to the nuclear case~\cite{Capella:1997yv,Armesto:2003fi}
do not correspond to any definite order in a
power expansion in $1/Q$ but perform a re-summation of all
powers\footnote{In~\cite{Frankfurt:2002kd,Frankfurt:2003zd} the discrepancy
between the data and the results of the model when evolved through DGLAP
to smaller values 
of $Q^2$ from that, $Q_0^2=4$ GeV$^2$ which is taken as initial value for
evolution and where the parametrization of data is used, is considered as
evidence of the existence of large power-suppressed contributions. On the
other hand, in other models for diffractive
data~\cite{Capella:1997yv,Armesto:2003fi} the presence of strong
$Q^2$-dependent terms is not required to describe the nuclear data at
low $Q^2$.}. Finally,
in these models the comparison with experimental data, when such comparison is
available, turns out to be reasonable, see Fig.~\ref{fig8}
- although the $Q^2$-dependence of the
nuclear ratios results
too smooth compared to data~\cite{Arneodo:1996ru}, which apparently
indicates the need of additional DGLAP evolution.

\subsection{High-density QCD}
\label{hdqcd}

High-density QCD - the domain of large gluon densities - has become a very
fashionable subject in the last fifteen years. It deals with the behaviour
of QCD at very large energies. Regarding the contents of this
review, it offers a definite theoretical
framework to compute shadowing corrections although the high-energy
approximations involved make its applicability to the present experimental
situation a subject of intense debate. The literature on this matter is vast
and I will not cite but a few papers, referring the reader to the
contributions in~\cite{cargese} and to the recent
reviews~\cite{Iancu:2003xm,McLerran:2005kk,Kovner:2005pe,Triantafyllopoulos:2005cn}
where
all the relevant references can be found.

\begin{figure}[htb]
\begin{center}
\vskip -1cm
\includegraphics[width=13cm]{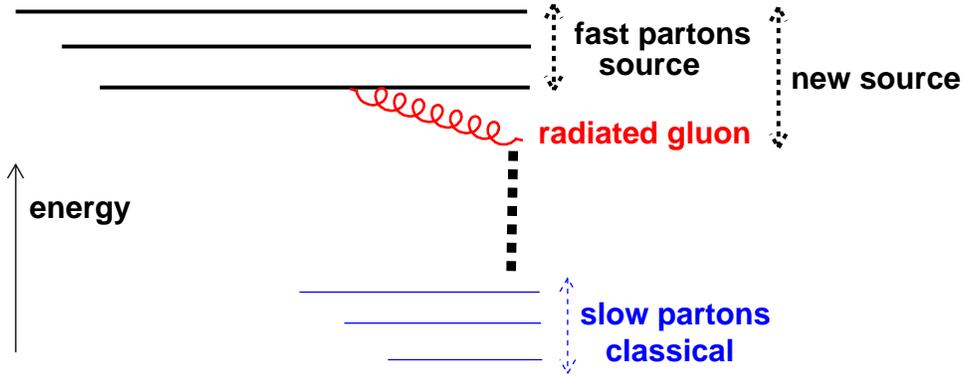}
\vskip -2cm
\caption{Diagram showing the separation between fast and slow partons, and the
contribution from radiated gluons.}
\label{fig9}
\end{center}
\end{figure}

In high-density QCD the small $x$ partons (slow gluons)
are treated classically due
to the high occupation number\footnote{For example, in the BFKL
framework~\cite{Kuraev:1977fs,Balitsky:1978ic} the gluon density $xg$
in the hadron
is expected to increase with decreasing $x$, $\propto x^{-2.65 \alpha_s}$.
The exponent in this power takes a value $\sim -0.5$ for the
strong coupling constant $\alpha_s\sim
0.2$. DIS proton data show
an increase $\sim x^{-0.3}$ for small $x$, although no conclusive
evidence of BFKL dynamics has been extracted from such behaviour.}
$\propto 1/\alpha_s$. This number is as high as it can be
- thus this field is often referred to as saturation
physics.
The source term for the classical
equations of motion comes from the fast partons e.g. valence quarks with
large $x$ (see Fig.~\ref{fig9}). The gluon density per unit impact
parameter and transverse momentum
of the gluon, the so-called unintegrated gluon density at fixed impact
parameter, computed in this way for an ultra-relativistic
large nucleus - the
McLerran-Venugopalan (MV)
model~\cite{McLerran:1993ni,McLerran:1993ka,McLerran:1994vd} - reads 
\begin{equation}
{dN^A_g\over dy \,d^2b \,d^2k_T}\equiv {d(xg_A)\over d^2b\,
d^2k_T} \propto
{1\over \alpha_s} \int {d^2x_T \over x_T^2}\, e^{-i x_T \cdot
k_T} \left(1-e^{-x_T^2 Q_s^2/4}\right).
\label{eq17}
\end{equation}
$Q_s^2\propto A T_A(b) xg_{\rm nucleon}$ is the squared saturation momentum or
saturation scale, which
corresponds to the typical gluon transverse momentum and to the scale at which
the exponential in (\ref{eq17})
starts to give large corrections. This saturation scale
increases with
increasing nuclear size and increasing energy or decreasing $x$. So it is
plausible that at some given high energy, $\alpha_s$ becomes small enough for
perturbative methods to be applied reliably.

\setcounter{footnote}{1}

With increasing energy, radiation processes start to contribute, see
Fig.~\ref{fig9}.
Additional
gluons are radiated from the source in a kinematical region
intermediate between the fast and
slow ones, and are absorbed in a redefinition of
the source. Mathematically this procedure results in
a renormalization group equation for the distribution of colour sources in the
hadron. Under several simplifications, this renormalization group equation
gives a single closed non-linear equation for the dipole-hadron scattering
amplitude\footnote{ See (\ref{eq11}), (\ref{eq12});
its relation with the unintegrated
gluon density comes through
the Bessel-Fourier transform defined in the right-hand side
of (\ref{eq17}).}, the
Balitsky-Kovchegov (BK) equation~\cite{Balitsky:1995ub,Kovchegov:1999yj}.
The use of the dipole model, see Subsection~\ref{glauber}, provides
its link with
the nuclear structure functions. Note that in the MV model, the number of
partons is not modified but they are only redistributed in transverse
momentum, while non-linear BK evolution does diminish the number of gluons.

From the explicit form of the MV model (\ref{eq17}), it is obvious that it
corresponds to a Glauber-like re-summation of rescatterings in the totally
coherent,
high-energy limit. Indeed, the MV
model can be used as initial condition at some not too small $x\sim 0.01$
for evolution towards smaller $x$ through the BK equation. Other initial
conditions
have been essayed in the literature.
But it is a
noticeable property of the BK equation that all initial conditions result in a
universal form for the solution of the equation after large enough
evolution~\cite{Armesto:2001fa,Lublinsky:2001bc}.
Such scaling implies, through the dipole model, that virtual
photon-nucleon cross sections are not a function of $x$, $Q^2$ and nuclear size
separately, but only of $Q^2/Q_s^2$ where all dependence on
$x$ and nuclear size is included in $Q_s$. This scaling has been found in
leptoproduction
data on nucleon and nuclear
targets~\cite{Stasto:2000er,Freund:2002ux,Armesto:2004ud}. Nevertheless the
situation is not yet clear: numerical
solutions~\cite{Albacete:2003iq,Albacete:2004gw}
of the BK equation show that
the asymptotic scaling behaviour appears at rapidities
larger than those presently available, which are still dominated by the
initial conditions. Besides the effects of a running coupling,
not included in the
derivation of the equation, are large.

\begin{figure}[htb]
\begin{center}
\includegraphics[width=7.8cm]{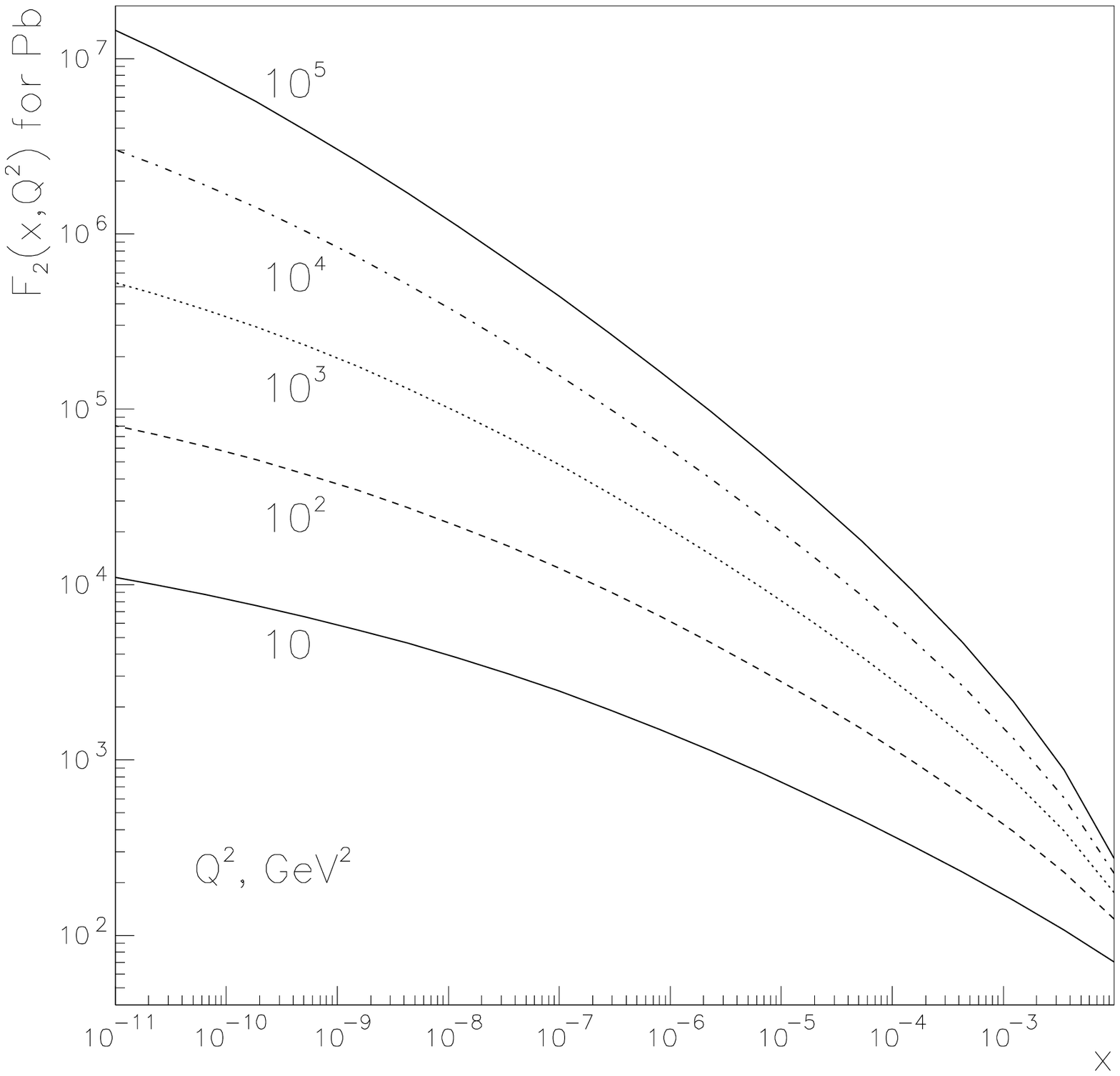}\includegraphics[width=7.8cm]{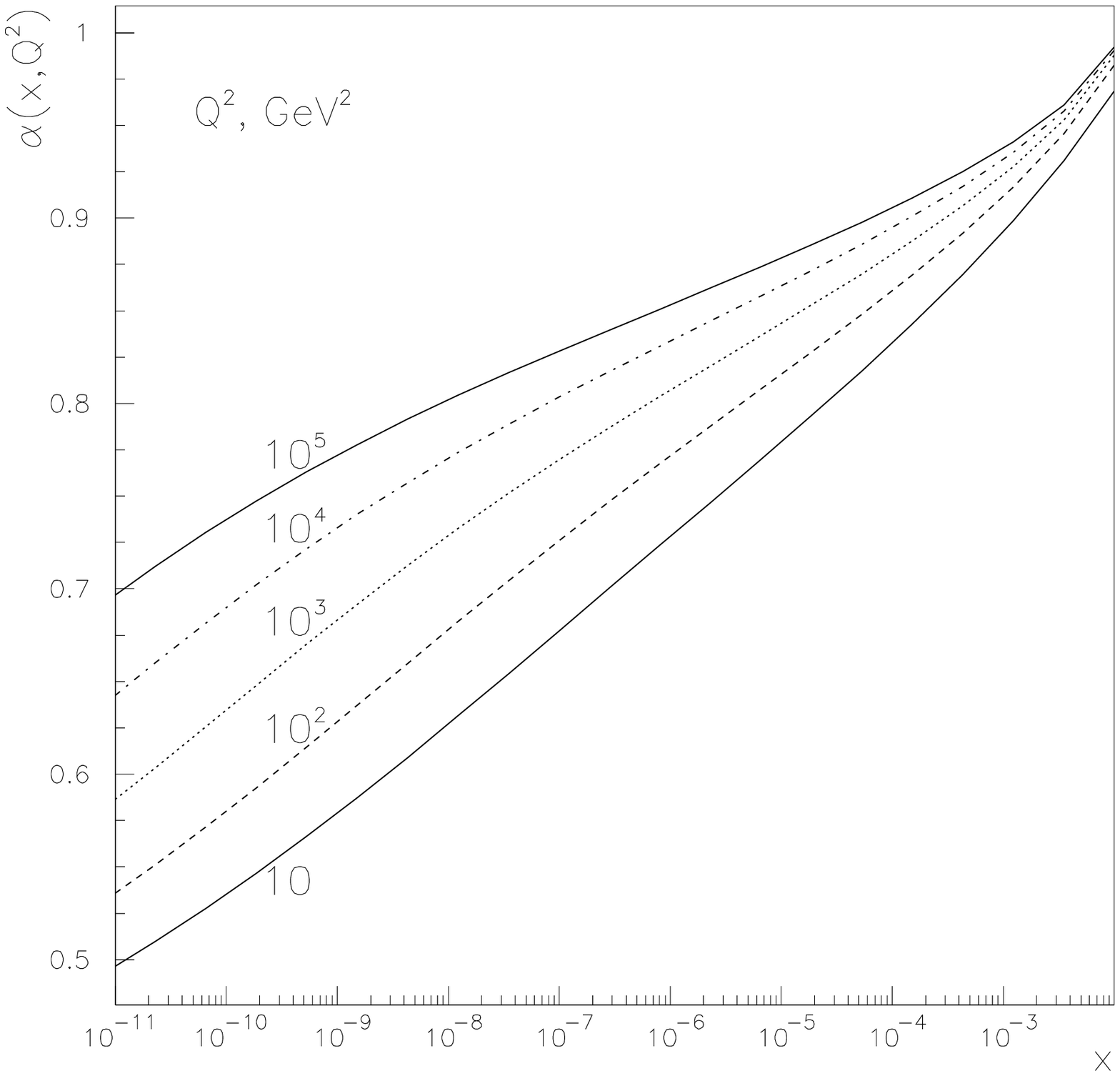}
\caption{Plot on the left: $F_2$ for Pb at small
$x$ for different $Q^2$ computed~\cite{Armesto:2001vm} through
the BK equation and the dipole model. Plot on
the right: nuclear dependence $\alpha$,
$F_2^A(x,Q^2)=A^{\alpha(x,Q^2)} F_2^{A=1}(x,Q^2)$ versus $x$ for different
$Q^2$.
[Figures taken from~\cite{Armesto:2001vm}.]}
\label{fig10}
\end{center}
\end{figure}

Predictions~\cite{Levin:2001et,Armesto:2001vm,Bartels:2003aa} for nuclear
structure functions of large nuclei at very small $x$ have been computed in
this framework. Some of them are presented in Fig.~\ref{fig10}.
As the main results in this approach, the nuclear structure function at very
small $x$ increases as $\ln^2{x}$ with decreasing $x$. It also
shows a $Q^2$-dependence
weaker than $\propto Q^2$
and a very strong nuclear dependence, with $\alpha(x,Q^2)$
($F_2^A(x,Q^2)=A^{\alpha(x,Q^2)} F_2^{A=1}(x,Q^2)$) getting
smaller than the geometrical 2/3 factor for very small values of
$x$~\footnote{Both this
very small value of $\alpha$ and one of the powers of $\ln x$ have their
origin in the
contribution of the dilute nuclear surface where the non-linear term in BK
evolution, or saturation effects in general, are negligible.}. Different
realizations~\cite{Levin:2001et,Armesto:2001vm,Bartels:2003aa} differ in the
initial conditions, in the consideration of impact parameter, and in the
treatment of large-size dipoles and of the evolution for not very small $x$.
They turn out to give results which may vary as much as a factor 10 for
$x\sim 10^{-7}$. Predictions for heavy flavour production also
exist~\cite{Armesto:2001vm,Goncalves:2003kp}.

The saturation scale computed within BK evolution behaves like
$Q_{s}^2\propto x^{-d \alpha_s}$, with $d=4\div 5$. Its dependence on the
nuclear size is not yet fully determined; in the most widely employed
approximation
valid for a very large nucleus, the $A$-dependence follows that of the
initial conditions, usually $\propto A^{1/3}$. Besides,
running coupling effects
modify both dependencies dramatically~\cite{Albacete:2004gw}. The saturation
scale can also be studied within phenomenological
approaches~\cite{Armesto:2002ny,Golec-Biernat:1998js,Stasto:2000er,Freund:2002ux,Armesto:2004ud}.
For example, a
value for the saturation scale can be obtained from Glauber approaches
(\ref{eq11}) as the value of $Q^2$ for which the effect of the
exponential factor in this
equation becomes
sizable (other geometrical criteria have also been essayed, like
percolation~\cite{Armesto:2000zh}).
Values extracted from this kind of studies are $Q_{s}^2
\sim A^{\delta} (x/0.01)^{-0.3}$
GeV$^2$, with $\delta \gtrsim 1/3$.

\begin{figure}[htb]
\begin{center}
\vskip 1.5cm
\includegraphics[width=9cm,height=14.5cm]{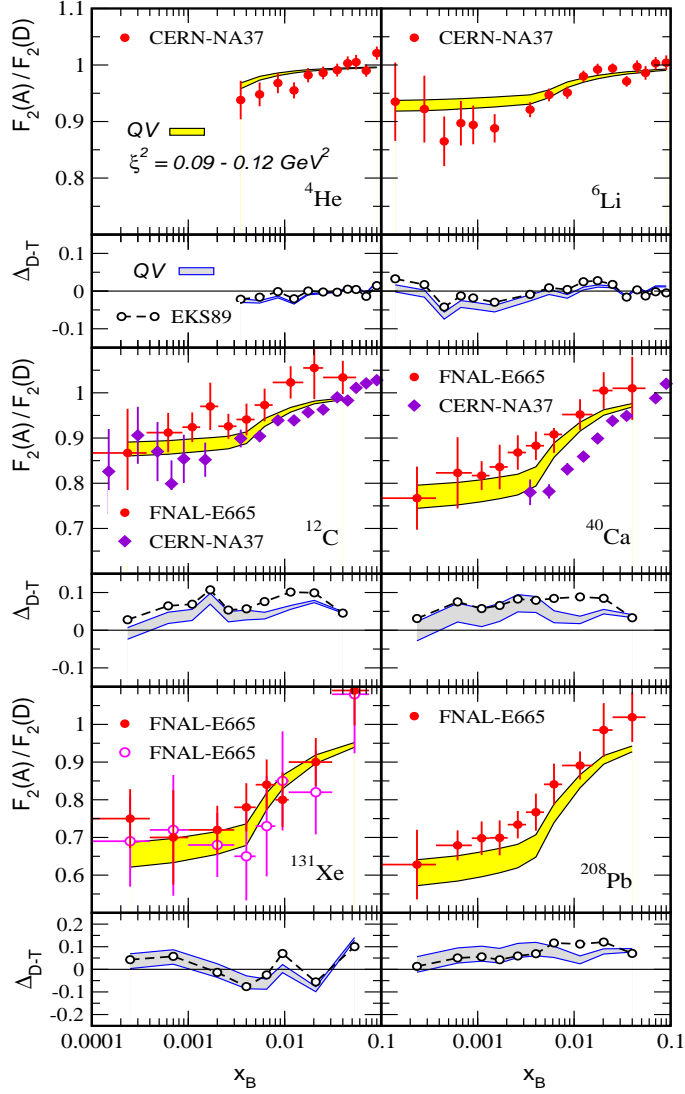}
\caption{$x$-dependence of the ratios in the model
in~\cite{Qiu:2003vd} for different nuclei,
compared with experimental
data~\cite{Amaudruz:1995tq,Arneodo:1995cs,Adams:1992nf,Adams:1995is} (filled
points).
The band corresponds to different choices of the scale of power corrections
$\xi^2$
in~\cite{Qiu:2003vd}.
$\Delta_{D-T} = {\rm Data}-{\rm Theory}$, and
the open circles joined by dashed lines in these plots
show the comparison to the
approach in~\cite{Eskola:1998df}.
The $(x,Q^2)$ correlation of the experimental points is taken into account
in the theoretical results shown here, as it was in those in Fig.~\ref{fig8}.
[Figure taken from~\cite{Qiu:2003vd}.]}
\label{fig11}
\end{center}
\end{figure}

Finally, other approach to the problem considers power-suppressed
corrections\footnote{The high-density QCD approach does not correspond to a
fixed order in the power expansion but re-sums, in some limit, all
power-suppressed contributions.} in $1/Q$. Such power-suppressed contributions
are enhanced by the nuclear size.
The first power-suppressed correction to DGLAP
evolution~\cite{Mueller:1985wy,Qiu:1986wh} results in a non-linear equation.
From the equality of the linear and non-linear terms, a value for the
saturation scale can be extracted~\cite{Accardi:2003be} which results in rough
agreement with the estimations previously discussed,
see the solid black
lines in Fig.~\ref{fig3}~\cite{Accardi:2003be}.
More recently,
such power-suppressed
contributions have been re-summed~\cite{Qiu:2003vd} in the high-energy eikonal
limit, resulting
in a rescaling of the $x$ variable whose results reasonably describe the
experimental data, see Fig.~\ref{fig11}. Besides they are in agreement with
available data on the nuclear effects on the longitudinal to transverse cross
sections~\cite{Abe:1998ym}.
Also more phenomenological studies~\cite{Castorina:2004ye}
are in agreement with the
experimental data.

\section{Models based on DGLAP evolution}
\label{dglap}

Another type of models do not try to address the origin of nuclear shadowing
(or of modifications of parton densities in nuclei in general) but to study the
$Q^2$-evolution of nuclear ratios of parton densities,
\begin{equation}
R_{i}^A(x,Q^2)={f_i^A(x,Q^2) \over A \, f_i^{\rm nucleon}(x,Q^2)}\,,\ \ f_i=q,
\bar q, g,
\label{eq18}
\end{equation}
through the DGLAP evolution
equations~\cite{Dokshitzer:1977sg,Gribov:1972ri,Altarelli:1977zs}, see
also~\cite{roberts}. From the
very first attempts~\cite{Eskola:1992zb}, several analysis have
appeared~\cite{Frankfurt:2002kd,Frankfurt:2003zd,Eskola:1998df,Indumathi:1996pb,Eskola:1998iy,Hirai:2001np,Hirai:2004wq,deFlorian:2003qf}.
They try to perform for the nuclear case the same program developed for the
nucleon: Nuclear ratios are parametrized at some value $Q_0^2\sim 1\div 2$
GeV$^2$ which is assumed large enough for perturbative DGLAP evolution to be
applied reliably. These initial parametrizations for
every parton density have to cover the full $x$ range $0<x<1$. In the
nuclear case, the nuclear size appears as an additional variable. Then these
initial conditions are evolved through the DGLAP equations towards larger
values of $Q^2$ and compared with experimental data. From this comparison the
initial parametrizations are adjusted.

Different approaches differ in several details,
see~\cite{Accardi:2003be}:
\begin{itemize}
\item The form of the parametrizations at the initial scale. For example,
in~\cite{Eskola:1998df,Eskola:1998iy} shadowing saturates for very small
$x$\footnote{The
model~\cite{AyalaFilho:2001cq}
uses Glauber-like rescatterings to effectively include some high-density
corrections to the parametrizations~\cite{Eskola:1998df,Eskola:1998iy}.
These corrections lead to a shadowing which does not saturate at small $x$,
and which increases with increasing
$Q^2$ for small $x\lesssim 10^{-4}$.},
contrary to~\cite{deFlorian:2003qf}.
Also the value of $Q_0^2$ varies e.g. from $\sim 0.4$
GeV$^2$~\cite{deFlorian:2003qf} to 2.25
GeV$^2$~\cite{Eskola:1998df,Eskola:1998iy}. The parametrizations for sea
quarks and gluons in~\cite{Hirai:2001np} do not show any EMC effect.
Special mention has to be done to
the approach of~\cite{Frankfurt:2002kd,Frankfurt:2003zd} where the initial
gluon density is taken from diffractive nucleon data at $Q_0^2=4$ GeV$^2$
as discussed in Subsection~\ref{gribov} and no attempt is made to modify it
from the comparison with experimental data.

\item The use of different sets of experimental data. For example, Drell-Yan
data~\cite{Alde:im} are used
in~\cite{Eskola:1998df,Eskola:1998iy,Hirai:2004wq,deFlorian:2003qf}
but not in~\cite{Hirai:2001np}. These data give the main constraint to the
valence and sea contributions in the antishadowing region
in~\cite{Eskola:1998df,Eskola:1998iy} but the parametrizations
in~\cite{Hirai:2004wq} do not show antishadowing for sea quarks.
HERMES data~\cite{Ackerstaff:1999ac} are used in~\cite{Hirai:2004wq}.
Also the data on
the $Q^2$-dependence of nuclear ratios~\cite{Arneodo:1996ru} are included
in~\cite{Eskola:1998df,Eskola:1998iy,Hirai:2004wq,deFlorian:2003qf} but not
in~\cite{Hirai:2001np}; they give the main constrain on the gluon distribution
at small and moderate $x$, see below.

\item The order of DGLAP evolution. The evolution is made at leading order (LO)
in~\cite{Eskola:1998df,Eskola:1998iy} and at next-to-leading (NLO) order in
~\cite{Hirai:2001np,Hirai:2004wq,deFlorian:2003qf}. This turns out to modify the
$Q^2$-dependence of nuclear ratios.

\item  The treatment of isospin effects and the use of sum rules as additional
constraints for evolution. For
example, isospin symmetry of the nuclear
ratios is assumed in~\cite{Eskola:1998df,Eskola:1998iy} but
not in~\cite{Hirai:2001np}.
Momentum, charge and baryon number conservation are used
in~\cite{Hirai:2001np,deFlorian:2003qf}, but charge conservation is not used
in~\cite{Eskola:1998df,Eskola:1998iy}. In practice, these differences result
numerically small.

\item The different nucleon partons densities used in the analysis. In
practice this choice is of little importance at the level of the nuclear ratios,
as its effect appears in both the numerator and denominator in (\ref{eq18})
and cancels to a large extent.
\end{itemize}

\begin{figure}[htb]
\begin{center}
\includegraphics[width=13cm]{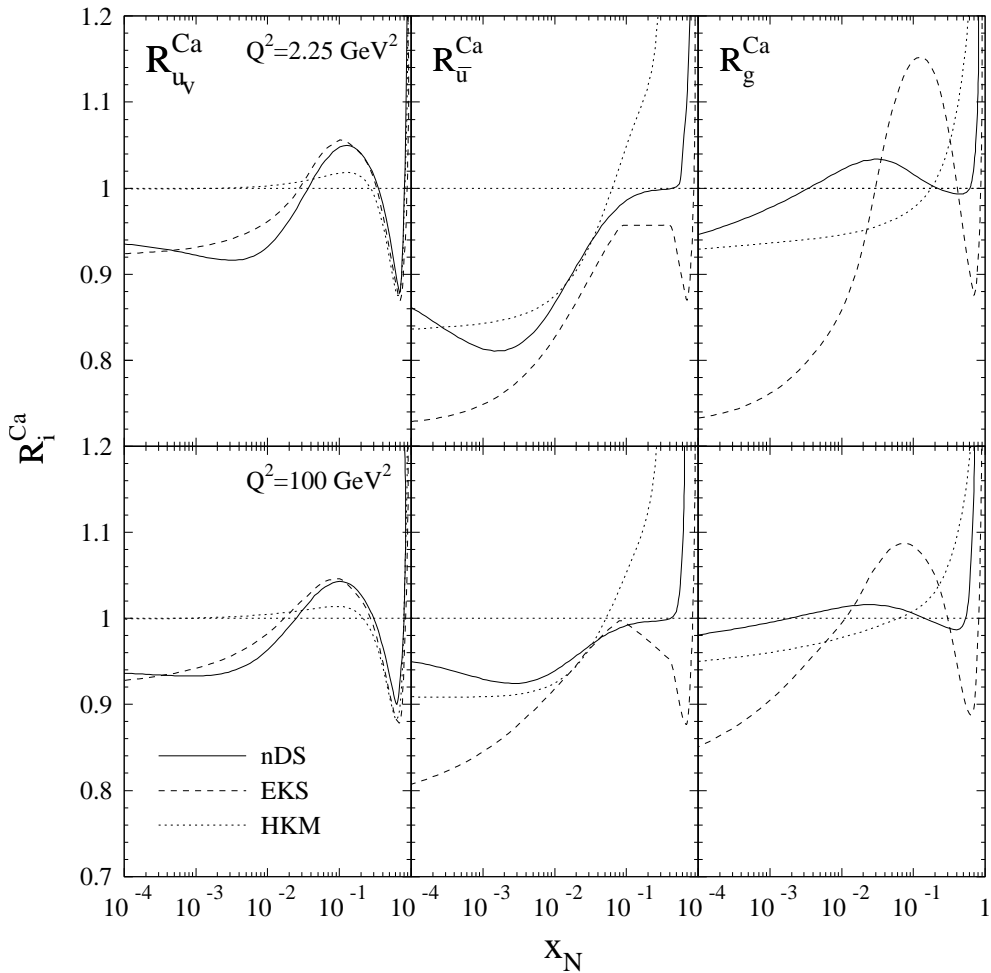}
\caption{Nuclear ratios for Ca versus $x$
computed for different partons densities, for
two different values of $Q^2$. Results
from~\cite{deFlorian:2003qf}
(solid),~\cite{Eskola:1998df} (dashed)
and~\cite{Hirai:2001np} (dotted lines) are compared.
[Figure taken from~\cite{deFlorian:2003qf}.]}
\label{fig12}
\end{center}
\end{figure}

A comparison of different approaches can be found in Fig.~\ref{fig12}; see
also Section~\ref{compa}. The differences are noticeable,
even more when one
considers that all approaches have been designed to
reproduce available experimental data (but see
below the discussion on the $Q^2$-dependence of nuclear ratios),
see~\cite{Accardi:2003be} for comments. Concerning shadowing, it turns
out that the nuclear ratios for gluons are almost unconstrained for $x<0.02$,
and for sea quarks for $x<0.005$. The stronger constraint on gluons in the
region $0.02<x<0.2$ comes from the $Q^2$-dependence of nuclear ratios which I
will discuss now.

\begin{figure}[htb]
\begin{center}
\includegraphics[width=13cm]{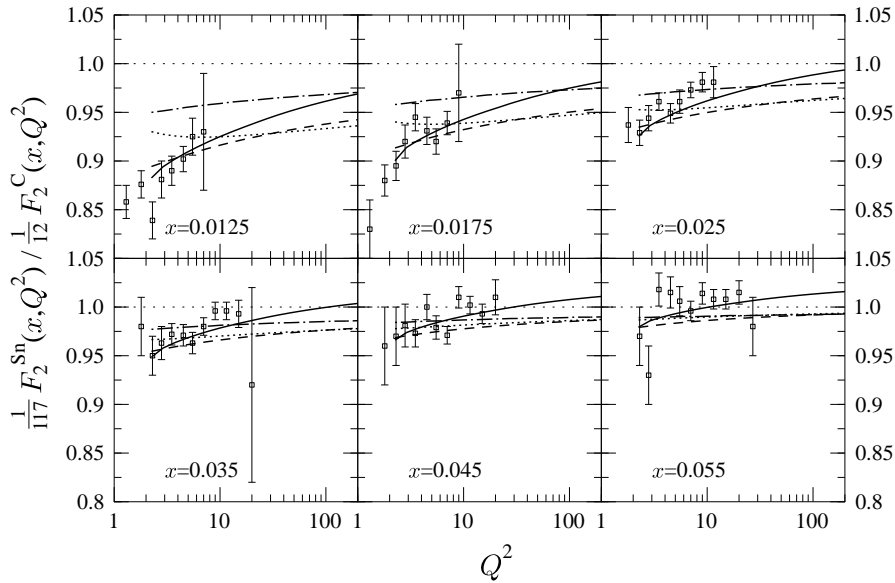}
\vskip -1cm
\caption{Results of the
models~\cite{Eskola:1998df} (solid)
and~\cite{Hirai:2001np} (dashed-dotted), and
of DGLAP evolution of initial conditions
from~\cite{hpc} (dashed)
and~\cite{Li:2001xa}
(dotted lines), for the $Q^2$-evolution of the ratio Sn/C at different values
of $x$, compared with the experimental data~\cite{Arneodo:1996ru}.
[Figure taken
from~\cite{Eskola:2002us}.]}
\label{fig13}
\end{center}
\end{figure}

The DGLAP equations establish a relation between the logarithmic
$Q^2$-evolution of the structure functions and the gluon
distribution~\cite{Prytz:1993vr}. Such relation, valid at LO and small $x$,
has been extended to the
nuclear ratios~\cite{Eskola:2002us}:
\begin{equation}
\frac{\partial R_{F_2}^A(x,Q^2)}{\partial \ln Q^2}
\approx
\frac{10\alpha_s}{27\pi}\frac{xg(2x,Q^2)}{\frac{1}{2}F_2^{\rm deuterium}(x,Q^2)}
\biggl\{R_g^A(2x,Q^2)-R_{F_2}^A(x,Q^2)\biggr\}.
\label{eq19}
\end{equation}
In this way, $R_g^A(2x,Q^2)>R_{F_2}^A(x,Q^2)$ implies a positive $Q^2$ slope,
while $R_g^A(2x,Q^2)<R_{F_2}^A(x,Q^2)$ gives a negative slope. The available
data on the $Q^2$-dependence of nuclear ratios~\cite{Arneodo:1996ru} allow to
constraint within the DGLAP evolution scheme, the relation between the nuclear
gluon distribution and the nuclear ratio for $F_2$.
In Fig.~\ref{fig13} the result of $Q^2$-evolution of the nuclear $F_2$-ratio
for Sn over C is shown~\cite{Eskola:2002us} for different
models~\cite{Eskola:1998df,Hirai:2001np}, and also for DGLAP-evolved
parametrizations~\cite{Li:2001xa,hpc} (these parametrizations were originally
proposed as $Q^2$-independent). While in the parametrization~\cite{hpc}
the ratio for all flavours is equal at the initial scale $Q_0^2$,
and thus it gives a too small but
positive slope, in the parametrization of ~\cite{Li:2001xa} the shadowing for
gluons is much larger than that for sea and valence quarks so it  results in a
negative slope at the smallest $x$ and $Q^2$.
Therefore, from the comparison with experimental
data~\cite{Arneodo:1996ru}, DGLAP analysis favours those sets in which gluons
are less shadowed than quarks for $x\sim 0.01$. This is at variance with some
approaches e.g.~\cite{Frankfurt:2002kd,Frankfurt:2003zd} where gluons are more
shadowed than quarks, see the next Section.
As commented in Subsection~\ref{gribov},
the discrepancy
between the data and the results of this model when evolved to smaller values
of $Q^2$ from $Q_0^2=4$ GeV$^2$
is considered as
evidence of the existence of large power-suppressed
contributions\footnote{DGLAP equations consider only leading power-suppressed
contributions - thus the shadowing produced by DGLAP evolution is referred to
as leading-twist shadowing.
Nevertheless attempts have been done to include the first
power-suppressed corrections~\cite{Mueller:1985wy,Qiu:1986wh}
in this analysis of the $Q^2$-dependence
of nuclear ratios~\cite{Eskola:1992zb,Eskola:2002us}.
But the result of such contributions
is to make the logarithmic slope smaller, so they do not help to solve this
problem.}.

While DGLAP approaches do not address the fundamental problem of the origin of
shadowing, they are of great practical interest. They provide the
parton densities required to compute cross sections for observables
characterized by a hard scale for which collinear
factorization~\cite{Collins:1989gx} can be applied,
see Section~\ref{appli} and~\cite{Accardi:2003be}
for discussions and further references. As a final comment, the centrality
dependence of shadowing is not addressed in these models as the existing
experimental data do not allow its determination, although some
approaches e.g.~\cite{Li:2001xa,Emel'yanov:1999bn}
provide an ansatz for such a dependence.

\section{Comparison among different models}
\label{compa}

\begin{figure}[htb]
\begin{center}
\includegraphics[width=13cm]{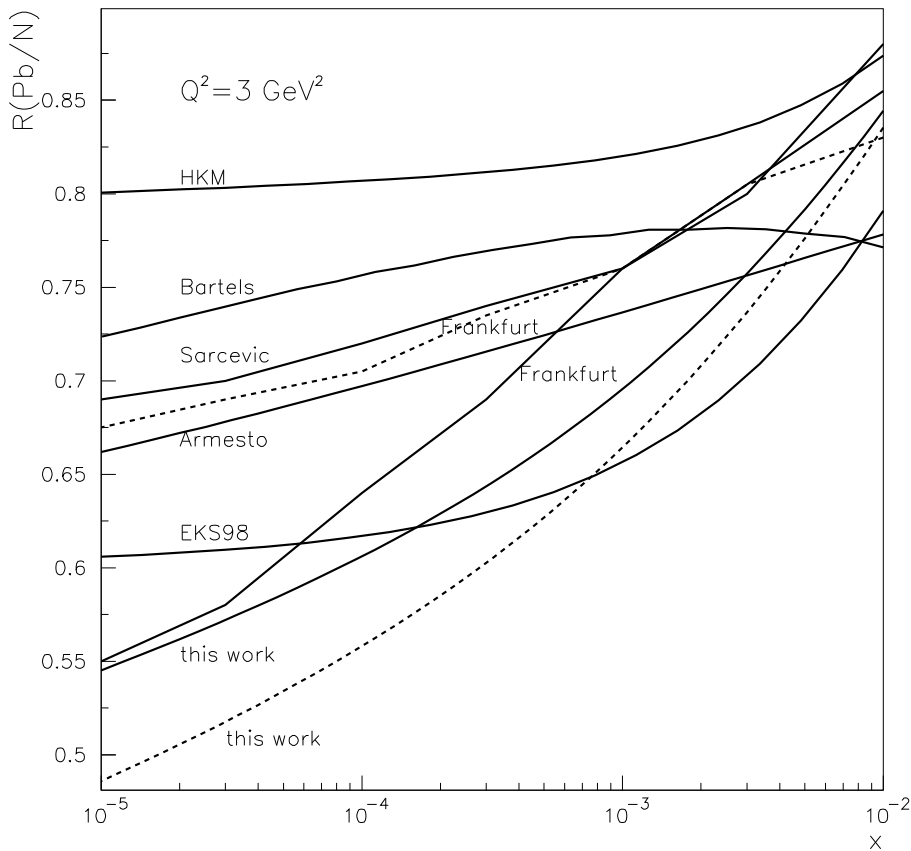}
\vskip -0.5cm
\caption{$F_2$-ratios for Pb
versus $x$ at fixed $Q^2=3$ GeV$^2$ from the models: Armesto {\it et al.}
(``this work")~\cite{Armesto:2003fi},
HKM~\cite{Hirai:2001np}, Sarcevic~\cite{Huang:1997ii},
Bartels~\cite{Bartels:2003aa},
Frankfurt~\cite{Frankfurt:2002kd} (at
$Q^2=4$ GeV$^2$),
Armesto~\cite{Armesto:2002ny} and EKS98~\cite{Eskola:1998df,Eskola:1998iy}.
[Figure taken
from~\cite{Armesto:2003fi}.]}
\label{fig14}
\end{center}
\end{figure}

In Fig.~\ref{fig14} a comparison of the results of different models for
$R_{F_2}^{\rm Pb}$ is shown. The results coincide within $\sim 15$ \% in the
region $x\sim 0.01$ where experimental data exists. But they strongly disagree
for smaller values of $x$, the difference being almost a factor 2 at $x\sim
10^{-5}$. In general, models~\cite{Frankfurt:2002kd,Armesto:2003fi}
based on Gribov inelastic shadowing give a larger shadowing than those based
in Glauber-like rescatterings~\cite{Huang:1997ii,Armesto:2002ny}. The
model~\cite{Bartels:2003aa} based on high-density QCD gives less shadowing.
Among the models based on DGLAP evolution,~\cite{Eskola:1998df,Eskola:1998iy}
gives larger shadowing than~\cite{Hirai:2001np} and to the one
in~\cite{deFlorian:2003qf}, see Fig.~\ref{fig12} for sea quarks which
determine $F_2$ for small $x$. But one
should take into account that in DGLAP approaches the small $x$ behaviour at
$Q^2$ close to $Q_0^2$
comes mainly from the assumptions for the initial conditions.
The model~\cite{AyalaFilho:2001cq} gives results for $R_{F_2}^{\rm Pb}$ at
$Q^2=2.25$ GeV$^2$
very close to those of~\cite{Eskola:1998df,Eskola:1998iy} at $x=0.01$ where
the considered high-density corrections are almost negligible, but
sizably smaller, $R_{F_2}^{\rm Pb} \simeq 0.43$, at $x=10^{-5}$.
These predictions could be checked in the EIC~\cite{Deshpande:2005wd}, where
values of $x\sim 3\cdot 10^{-4}$ at $Q^2\sim 1$ GeV$^2$ should become
accessible, or in ultra-peripheral proton-ion or ion-ion
collisions (UPC)~\cite{Bertulani:2005ru,Goncalves:2005ge}.

Now let me turn to the nuclear gluon density.
In those approaches~\cite{Huang:1997ii,Armesto:2002ny} which rely on the dipole
model,
the gluon density is obtained from the unintegrated gluon
distribution through
\begin{equation}
xg_A(x,Q^2)=\int_{\Lambda^2}^{Q^2} d^2k_T\int d^2b\, {d(xg_A)\over d^2b\,
d^2k_T}\,,
\label{eq20}
\end{equation}
where $\Lambda^2$ is some infrared cut-off, if required.
This identification is only true
for $Q^2\gg Q_{\rm
sat}^2$~\cite{Kovchegov:1998bi,Mueller:1999wm,Collins:2003fm}.
In models based on Gribov inelastic
shadowing~\cite{Frankfurt:2002kd,Tywoniuk:2005df}, the nuclear gluon density
is obtained from (\ref{eq13})-(\ref{eq16}) but using the gluon density in the
pomeron instead of the diffractive structure function of the proton.

\begin{figure}[htb]
\begin{center}
\includegraphics[width=13cm]{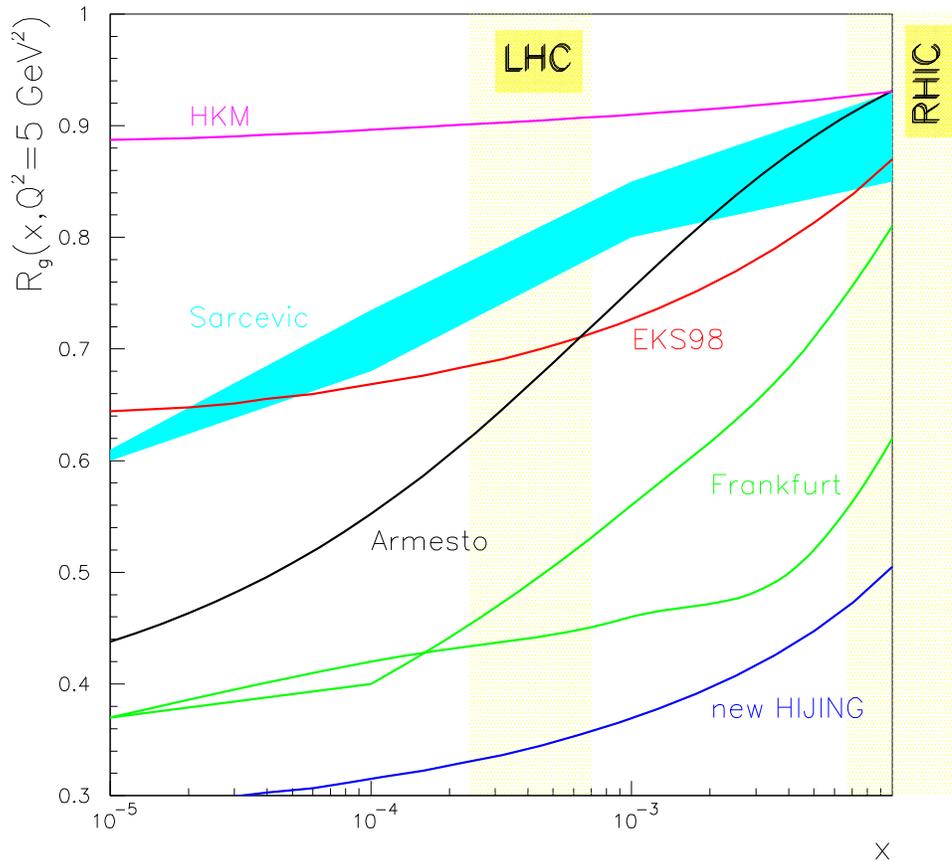}
\vskip -0.5cm
\caption{Ratios of gluon distribution functions for Pb versus $x$
from different models at
$Q^2=5$ GeV$^2$: HKM~\cite{Hirai:2001np},
Sarcevic~\cite{Huang:1997ii}, EKS98~\cite{Eskola:1998df,Eskola:1998iy},
Frankfurt~\cite{Frankfurt:2002kd},
Armesto~\cite{Armesto:2002ny} and
new HIJING~\cite{Li:2001xa}. RHIC and LHC point the ranges of
$x=(Q/\sqrt{s_{_{\ensuremath{\it{NN}}}}})e^y$ for processes with $|y|\le0.5$, $Q^2=5$ GeV$^2$ at RHIC
($\sqrt {s_{_{\ensuremath{\it{NN}}}}})
=200$ GeV) and LHC ($\sqrt {s_{_{\ensuremath{\it{NN}}}}})=
5.5$~TeV) respectively.
[Figure taken
from~\cite{Accardi:2003be}.]}
\label{fig15}
\end{center}
\end{figure}

In  Fig.~\ref{fig15} a comparison of the results of different models for
$R_{g}^{\rm Pb}$ is shown. At variance to the case of $R_{F_2}^{\rm Pb}$, now
there are large discrepancies also at $x\sim 0.01$, see the discussion on the
$Q^2$-dependence of the nuclear ratios at the
end of Section~\ref{dglap}. This is due to the fact that the gluon density is
only indirectly constrained by experimental data - $F_2$ is mainly determined
by the sea quarks at such values of $x$. The discrepancy between the different
approaches is roughly a factor 2 at $x\sim 0.01$ and a factor 3 at $x\sim
10^{-5}$. Among the DGLAP approaches the results
of~\cite{Li:2001xa}, in which gluon shadowing is fixed to
reproduce the multiplicity in Au-Au collisions at RHIC (see the next
Section), are
clearly below those from~\cite{Eskola:1998df,Eskola:1998iy}
and~\cite{Hirai:2001np}\footnote{This model gives results very close to those
from~\cite{deFlorian:2003qf}, see Fig.~\ref{fig12} for gluons, although the
latter shows an increase towards one at very small $x$ due to the form of the
initial parametrizations.}, as discussed at the
end of Section~\ref{dglap}. Again, these DGLAP
results
are mainly determined by the initial parametrizations. The results of
Glauber-like models are similar at $x\sim 10^{-5}$ but those
of~\cite{Armesto:2002ny} are smaller than the ones in~\cite{Huang:1997ii} at
$x\sim 0.01$. Results from Gribov inelastic
shadowing~\cite{Frankfurt:2002kd} show a strong
shadowing: see Fig.~\ref{fig15}
and the mentioned discussion in Section~\ref{dglap};~\cite{Tywoniuk:2005df}
gives $R_{g}^{\rm Pb}\sim 0.7$ at $x=0.01$ and $\sim 0.2$ at $x=10^{-5}$ for
$Q^2=6.5$ GeV$^2$. The model~\cite{AyalaFilho:2001cq} gives results for
$R_{g}^{\rm Pb}$ at
$Q^2=2.25$ GeV$^2$
very close to those of~\cite{Eskola:1998df,Eskola:1998iy} at $x=0.01$ - as it
was the case for $R_{F_2}$, but
sizably smaller, $R_{F_2}^{\rm Pb} \simeq 0.37$, at $x=10^{-5}$.
Finally,~\cite{Kopeliovich:1999am} gives $R_{g}^{\rm Pb}\sim 0.95$ at $x=0.01$
and $\sim 0.75$ at $x=10^{-4}$ for
$Q^2=4$ GeV$^2$.

Contrary to $F_2$ which is directly measurable,
the gluon distribution is only indirectly constrained
both at an EIC~\cite{Deshpande:2005wd}, in proton-nucleus or nucleus-nucleus
collisions at RHIC~\cite{Adcox:2004mh,Back:2004je,Arsene:2004fa,Adams:2005dq}
and the LHC~\cite{Accardi:2003be}, and in
UPC~\cite{Bertulani:2005ru,Goncalves:2005ge}.
Some of these indirect
constraints in proton-nucleus
and
nucleus-nucleus
collisions
will be discussed in the next Section.

\section{Shadowing and particle production in high-energy
nuclear collisions}
\label{appli}

In this Section I will discuss some consequences of the phenomenon of shadowing
in high-energy nuclear collisions. I will devote most of the Section to
proton(deuteron)-nucleus collisions, and only at the end I will briefly refer
to nucleus-nucleus collisions.

Particle production is related to the parton densities in the projectile and
target through a factorization theorem, which in general takes the schematic
form
\begin{equation}
\sigma_{A-B\to CX}\propto \phi_{i/A}(x_A) \otimes \phi_{j/B}(x_B)
\otimes \hat{\sigma}_{ij \to C}(x_A,x_B,Q_C).
\label{eq21}
\end{equation}
In this formula, integration over the relevant variable is implicit.
$\hat{\sigma}$ is a scattering matrix for partons $i,j$ to
give the produced parton/particle $C$.
$Q_C$ is a scale (e.g. mass, transverse momentum,$\dots$) characteristic of
$C$.
And $\phi_{i/A}(x_A)$ are the probabilities of finding parton $i$ in
hadron/nucleus $A$ with momentum fraction $x_A$. Depending on the type of
factorization, collinear for $\Lambda_{\rm QCD} \ll Q_C \ll
\sqrt{s}$~\cite{Collins:1989gx} or
$k_T$ for $\Lambda_{\rm QCD} \ll Q_C \lesssim
\sqrt{s}$~\cite{Catani:1990eg,Collins:1991ty,Levin:1991ry}, the matrix elements
are different and the partons distributions are integrated or unintegrated
respectively~\cite{Collins:2003fm}.
For the status of collinear factorization in collisions
involving nuclei, see~\cite{Accardi:2003be}. The validity and exact
form
of $k_T$
factorization in proton-nucleus collisions - here proton means a hadron
in which the parton density is small - is under discussion, see
e.g.~\cite{Braun:2006wj,Kovchegov:2001sc,Nikolaev:2005qs}
and references therein\footnote{In~\cite{Nikolaev:2000sh,Nikolaev:2004cu},
a non-linear $k_T$ factorization
is proposed and the properties of the resulting nuclear gluon density are
studied.}. For
nucleus-nucleus~\cite{Braun:2000bh,Kharzeev:2000ph,Kharzeev:2001gp},
arguments exist which indicate that it is not
valid~\cite{Balitsky:2004rr,Fujii:2005vj}, but the size of the violations has
not been fully quantified yet.
Therefore, most of the discussions in the literature
assume the validity of such
factorizations in the nuclear case, which will be done in the following.

In a $2\to 2$ process the relation between
rapidity $y$ and transverse mass $m_T=\sqrt{m^2+p_T^2}$ of the final
partons/particles and their fractional momenta $x_{A,B}$ is
\begin{equation}
x_{A,B}={m_T\over \sqrt{s}} \, \exp{(\pm y)}.
\label{eq22}
\end{equation}
At midrapidity at RHIC,
only particle production with small $p_T\lesssim 1\div 2$ GeV will
be sensitive to the shadowing region in parton densities. At the LHC, the
region of transverse momenta will be much larger, see Fig.~\ref{fig3} for the
relevant kinematical regions which will be studied in proton-nucleus
collisions at RHIC and the LHC for different observables.
In~\cite{Accardi:2003be} extensive studies of the impact of parton densities
on particle production with a large scale (e.g. transverse momentum or mass)
at the LHC can be found.

Most of the discussions at RHIC are done in terms of the so-called nuclear
modification factors:
\begin{equation}
R_{A-B}={{\rm produced\ in}\ A-B\over {\rm expected\ in}\ A-B} = {N_{A-B}\over
N_{coll}\times N_{\rm nucleon-nucleon}}\,.
\label{eq23}
\end{equation}
Here $N_{A-B}$ is the number of produced particles of a given type
in some kinematical region,
in $A$-$B$ collisions.
$N_{coll}$ is the number of binary incoherent
nucleon-nucleon collisions with which
particle production is expected to scale in collinear factorization, so
$R_{A-B}\to 1$ for a large scale of the produced particle.

\begin{figure}[htb]
\begin{center}
\includegraphics[width=12cm]{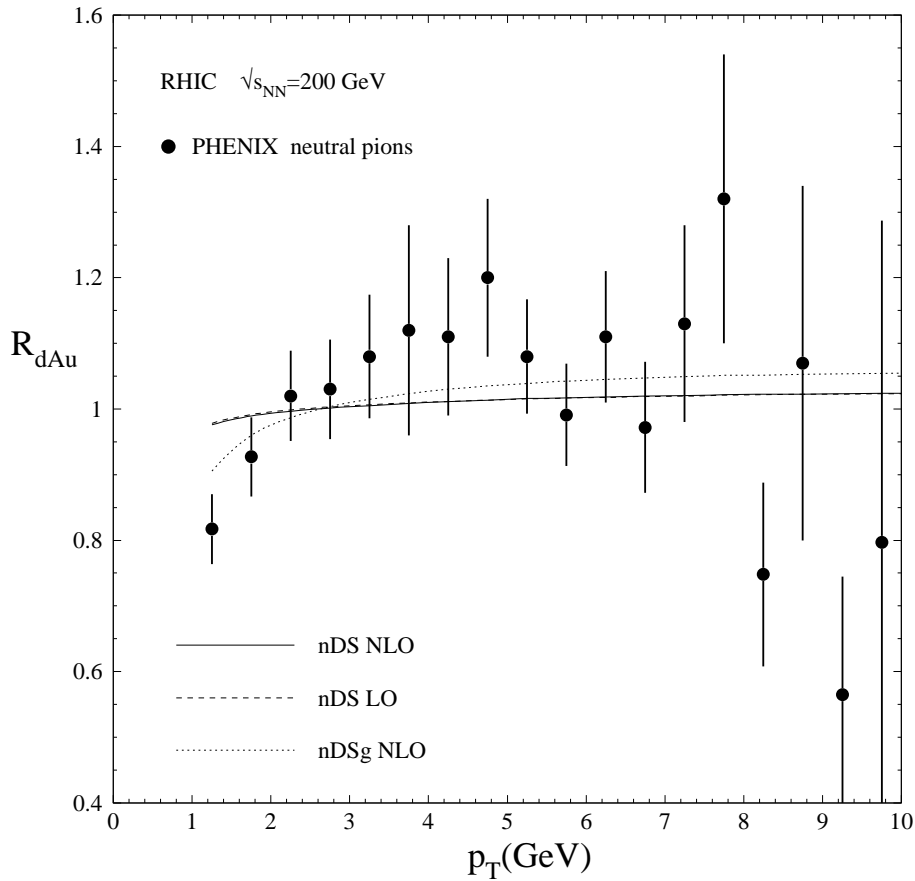}
\vskip 0.5cm
\caption{Nuclear modification factor for neutral pions at $\eta\simeq 0$
in d-Au collisions at $\sqrt{s_{_{\ensuremath{\it{NN}}}}}=200$
GeV. The curves are the results from~\cite{deFlorian:2003qf} and the
experimental data from~\cite{Adler:2003ii}.
[Figure taken
from~\cite{deFlorian:2003qf}.]}
\label{fig16}
\end{center}
\end{figure}

In Fig.~\ref{fig16} the nuclear modification factor
for neutral pion production in d-Au
collisions at RHIC for pseudorapidity $\eta\simeq 0$
is shown and compared with results from collinear
factorization~\cite{deFlorian:2003qf}. Nuclear effects in deuterium,
as stated
previously, are usually neglected and, in any case, very small.
From this Figure it becomes clear that at midrapidities only the region of small
transverse momentum ($p_T\lesssim 1$ GeV)
is sensitive to the amount of shadowing in parton
densities.
Above $p_T\sim 3$ GeV, one can see a (small) enhancement
of particle production in nuclear targets - a phenomenon known as
the Cronin effect for more than 20 years~\cite{Cronin:1974zm}.

\begin{figure}[htb]
\begin{center}
\includegraphics[width=13cm]{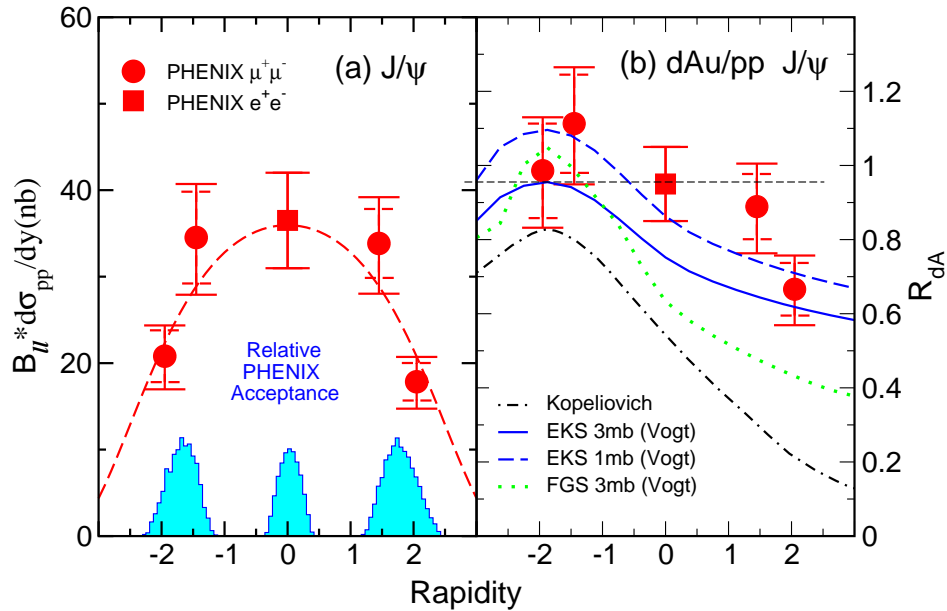}
\vskip -0.5cm
\caption{Plot on the left:
rapidity distribution of the J/$\psi$ cross section. Plot on the right:
nuclear modification factor for J/$\psi$~\cite{Adler:2005ph} in d-Au
collisions at $\sqrt{s_{_{\ensuremath{\it{NN}}}}}=200$
GeV. The curves in
this plot labeled Vogt~\cite{Vogt:2004dh}
correspond to different nuclear
absorption cross sections and different models for nuclear parton densities
EKS~\cite{Eskola:1998df,Eskola:1998iy} and FGS~\cite{Frankfurt:2002kd}, while
that labeled Kopeliovich corresponds to~\cite{Kopeliovich:2001ee}.
[Figure taken
from~\cite{Adler:2005ph}.]}
\label{fig17}
\end{center}
\end{figure}

A large quark mass also provides a scale which gives some justification to the
use of collinear factorization. Thus heavy flavour production can be employed
to constrain parton densities
inside nuclei~\cite{Armesto:1995zt,Armesto:1996xq} and their impact parameter
dependence~\cite{Klein:2003dj}.
At RHIC, direct measurements of open heavy flavour
production are still at the beginning~\cite{Tai:2004bf}. Most of the existing
data are obtained from electrons from semi-leptonic
decays~\cite{Adler:2005xv,Bielcik:2005wu} which are weakly correlated with the
parent heavy flavour. On the other hand, there exist studies of charmonium
production.
In Fig.~\ref{fig17} results for J/$\psi$
production in d-Au
collisions at RHIC~\cite{Adler:2005ph}
for different pseudorapidities are shown and compared~\cite{Vogt:2004dh} to
models for partons densities, using collinear factorization.
Such models imply
for forward
rapidities, see (\ref{eq22}), an additional suppression (i.e.
gluon shadowing as heavy flavour production is mostly sensitive to the gluon
channel) on top of
the nuclear absorption of quarkonium in
nuclei~\cite{Capella:1988ha,Gerschel:1988wn}.
From Fig.~\ref{fig17} right, models which do not consider a very strong gluon
shadowing~\cite{Eskola:1998df,Eskola:1998iy} are apparently
favoured
over those which
consider a large shadowing~\cite{Frankfurt:2002kd}, see
also the discussion at the end of Section~\ref{dglap}. But the large error
bars in the experimental data and the remaining
uncertainty in the amount of nuclear
absorption and its behaviour with
rapidity~\cite{Braun:1997qw,Kopeliovich:2001ee},
make it difficult to extract any definite conclusion.

\begin{figure}[htb]
\begin{center}
\includegraphics[width=15.7cm]{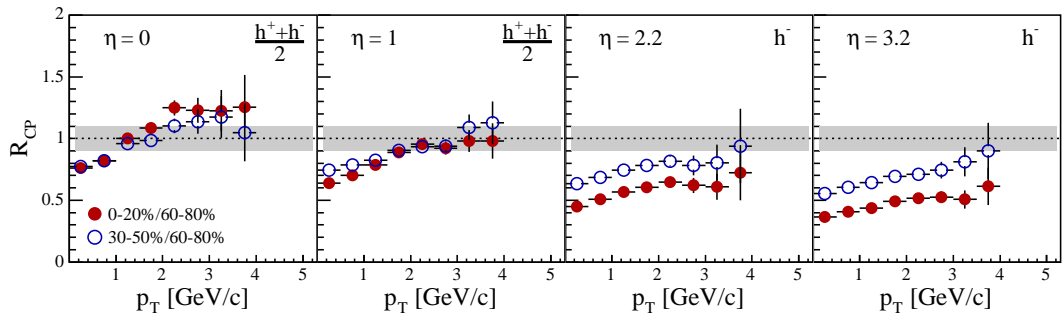}
\vskip 0.cm
\caption{Ratio of nuclear modification factors for different centrality classes,
for charged (two leftmost plots) and negative (two
rightmost plots)
particles at different pseudorapidities for central/peripheral
(filled points) and
semicentral/peripheral (open points) d-Au collisions at
$\sqrt{s_{_{\ensuremath{\it{NN}}}}}=200$
GeV~\cite{Arsene:2004ux}.
[Figure taken
from~\cite{Arsene:2004ux}.]}
\label{fig18}
\end{center}
\end{figure}

\begin{figure}[htb]
\begin{center}
\includegraphics[width=11cm]{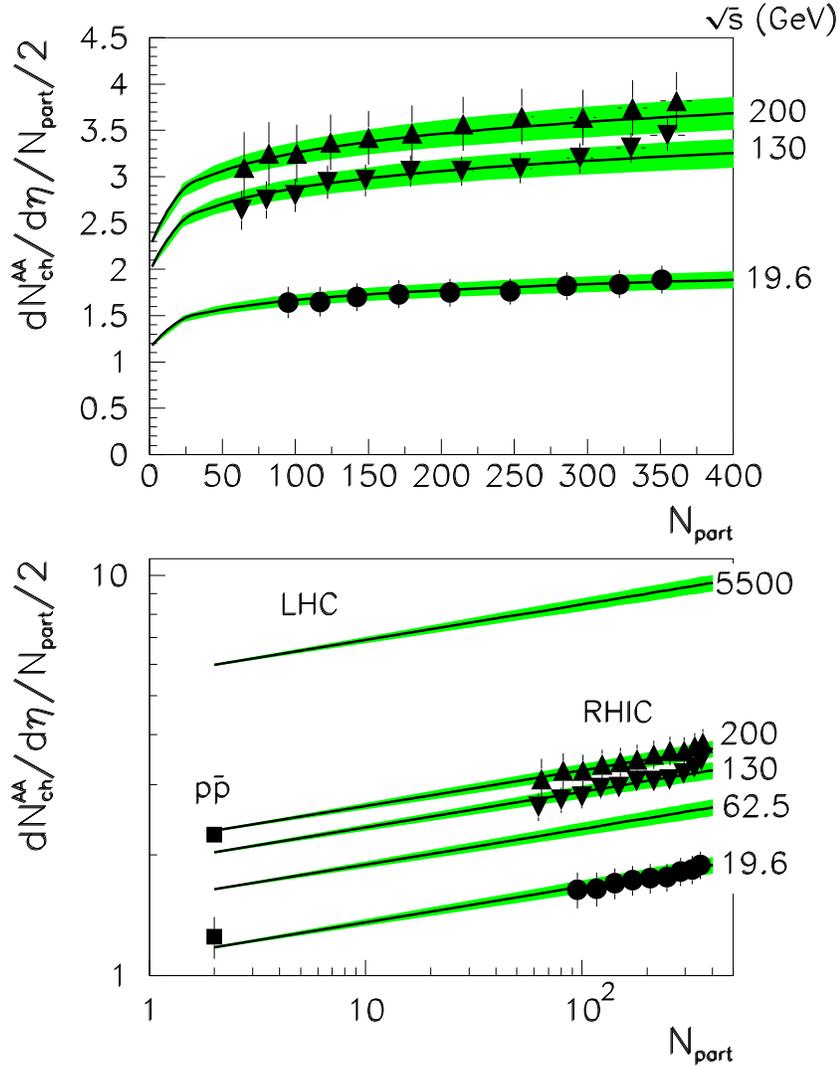}
\vskip -0.cm
\caption{Energy and centrality dependence (in terms of the number $N_{\rm part}$
of nucleons
participating in the collision) of the multiplicity of charged
particles at midrapidity divided by $N_{\rm part}$,
in Au-Au collisions at several RHIC energies and for LHC energies. The
results~\cite{Armesto:2004ud} of $(2/N_{\rm part})
[dN^{A-A}_{\rm ch}/d\eta](\eta\sim 0)=0.47[\sqrt{s_{_{\ensuremath{\it{NN}}}}}]
^{0.288}
N_{\rm part}^{0.089}$ are compared to PHOBOS data
\cite{Back:2002uc,Back:2004dy}.
Also shown in the lower panel are the ${\bar p}$-$p$ data
\cite{Thome:1977ky,Alner:1986xu}, as quoted in
\cite{Back:2004dy}.
[Figure taken
from~\cite{Armesto:2004ud}.]}
\label{fig19}
\end{center}
\end{figure}

A striking finding in d-Au collisions at RHIC has been the change from an
enhancement of the nuclear modification factor for light particles at central
pseudorapidities (the Cronin effect, see Fig.~\ref{fig16}) to
a suppression, at all measured transverse momenta, at forward
pseudorapidities~\cite{Arsene:2004ux,Adler:2004eh,Back:2004bq,Adams:2006uz}
(Fig.~\ref{fig18})\footnote{The decrease of the nuclear modification factors
for light and heavy hadrons with increasing rapidity has been a well known
phenomenon for many years, but the transverse momentum
structure of the suppression far
from midrapidity has only recently been measured.},
see the review~\cite{Jalilian-Marian:2005jf}.
\setcounter{footnote}{1}
In Glauber-like models as the MV model, the Cronin
effect results naturally from the transverse momentum broadening due to
multiple scattering. But it has been taken as a great success of saturation
physics that non-linear small-$x$ evolution was able to predict the suppression
at all transverse momentum from initial conditions which contained the Cronin
effect~\cite{Albacete:2003iq,Kharzeev:2002pc,Baier:2003hr,Kharzeev:2003wz}.
Studies within
collinear factorization~\cite{Guzey:2004zp} indicate that shadowing has to be
stronger than usually considered
in DGLAP or Gribov inelastic shadowing~\cite{Capella:2004xj} approaches,
in order to justify these forward data,
and that
other effects are at work. In any case the clear conclusion can be drawn that
such data require a large amount of shadowing, even more considering that
the average values of $x$ probed in the nucleus are not very small,
$\langle x \rangle\sim
0.01$ (assuming $2\to 2$ processes)~\cite{Guzey:2004zp}\footnote{This
value of $x$ lies at the upper border of applicability of saturation ideas, as
commented in Subsection~\ref{hdqcd}. Thus
it is not clear that they can be employed to explain this phenomenon.}.
Other observables like Drell-Yan
production~\cite{Accardi:2003be,Kopeliovich:2001hf,Gelis:2002fw,Baier:2004tj,Betemps:2004xr},
and midrapidity production of high-$p_T$
hadrons in
proton-nucleus collisions at
the LHC, should further clarify this issue.

d-Au collisions at RHIC have been crucial to establish the baseline on
top of which final state effects due to
a new state of deconfined matter, the Quark Gluon Plasma, can be
searched for and eventually characterized~\cite{Gyulassy:2004zy}.
Due to this possibility, the
determination of nuclear shadowing in nucleus-nucleus collisions is far more
complex than in proton-nucleus,
as many other effects can be at work. In any case, the influence of
shadowing on bulk particle production will be large, resulting in a strong
reduction of multiplicities in Au-Au collisions at RHIC and Pb-Pb collisions
at the LHC, see e.g. the reviews~\cite{Armesto:2000xh,Armesto:2004sa},
or~\cite{Kharzeev:2000ph,Kharzeev:2001gp,Braun:2006dk}.
For example, in the
approach~\cite{Armesto:2003fi} based on Gribov inelastic shadowing and
assuming some kind of factorization based on the AGK rules, reduction
factors in head-on heavy ion collisions
$\sim 2$ at RHIC and $\sim 4$ at the LHC can be expected for particle
production at mid-rapidity, compared to $p$-$p$
collisions scaled by the number of
binary collisions.
Other approaches~\cite{Armesto:2004ud}
based on saturation ideas which directly relate
the shadowing
measured in lepton-nucleus collisions to multiplicities in proton-nucleus
and nucleus-nucleus collisions, also predict large suppressions and result in
agreement with experimental data, see
Fig.~\ref{fig19}.
Nevertheless it must be stressed that the
uncertainties on the separation of effects in the nuclear wave functions from
other effects due to the collision are large, as there is no well
established factorization scheme presently
available for nucleus-nucleus collisions.

\section{Conclusions}
\label{conclu}

The phenomenon of shadowing is of large importance from a theoretical
point of view: the behaviour of the nuclear wave function at high energies
provides useful information for our understanding of QCD in such regime.
It has also very strong practical implications:
parton densities in nuclei are required to
predict and understand particle production in collisions involving nuclei.
In this article the phenomenon of nuclear shadowing has been introduced.
I have
discussed multiple scattering as the underlying physical mechanism
illustrated through a simple example of two scatterings, and
presented several models which make use of it. Then I have analyzed the DGLAP
approach which does not address the origin of shadowing but only the
evolution of nuclear parton densities
through the DGLAP equations. Next I have shown a comparison of the different
models for $F_2$ and $xg$. Finally, I have reviewed the
results in nuclear collisions for which
shadowing is expected to play a large role, with special emphasis on
proton(deuteron)-nucleus collisions at RHIC.

The uncertainties on gluon shadowing at small $x\lesssim 10^{-2}$
are as large as a
factor 3, see Fig.~\ref{fig15}. Multiple
scattering approaches suggest a large amount of shadowing and tend to indicate
that shadowing for gluons is stronger than that for quarks. But DGLAP
evolution disfavours such situation in view of the existing data on
$Q^2$-evolution of the nuclear
modification factors, at least for $x\sim 0.01$ where experimental data
lie. In this $x$-region the uncertainties in multiple scattering
approaches are large due to finite coherence length effects.
Besides, at the existing
values of $Q^2$ for small $x$ the validity of pure
DGLAP evolution or the need of power-suppressed corrections, re-summed in the
totally coherent limit in some approaches like high-density QCD,
is not yet clear. 
So more data at smaller $x$, eventually coming from a large energy
lepton-nucleus collider like the EIC~\cite{Deshpande:2005wd}, or
from UPC~\cite{Bertulani:2005ru,Goncalves:2005ge}, are needed.

d-Au collisions at RHIC provide very useful information if the
validity of some form of factorization is assumed. J/$\psi$ production data in
the forward (deuteron)
direction constrain, within collinear factorization, shadowing for
gluons, apparently favouring models with not very large shadowing. On the
other hand, nuclear
modification factors for charged or negative particles in the forward
direction show a suppression for all transverse momenta first predicted in the
framework of high-density QCD. They seem to
indicate that shadowing has to be larger than in most available
models. Both sets of experimental data truly imply more physical mechanisms
than nuclear shadowing, and some sizable uncertainty still comes from the use of
any kind of factorization.

Nucleus-nucleus collisions at RHIC and the LHC will be strongly affected by
shadowing.
Thus, the quantification of shadowing effects will be crucial for the eventual
characterization of a
dense medium where many additional physical processes may be at work.
In this respect, in
the near future the LHC will open new possibilities, both with the heavy ion
program and, mainly, with proton-nucleus runs~\cite{Accardi:2003be}.

\ack
It is a pleasure to thank the people with whom I have
collaborated on this subject through the last fifteen years:
J. L. Albacete, J. Bartels,
M. A. Braun, M. Cacciari,
A. Capella, A. Dainese,
E. G. Ferreiro,
A. B. Kaidalov,
A. Kovner,
J. G. Milhano, C. Pajares, C. A. Salgado, Yu. M. Shabelski
and U. A. Wiedemann.
I specially thank
J. \'Alvarez Mu\~niz, D. d'Enterria,
C. Pajares, C. A. Salgado, Yu. M. Shabelski and R. A. V\'azquez
for most useful
discussions, suggestions and a critical reading of this manuscript, and V. P.
Gon\c calves, V. Guzey and N. N.
Nikolaev for useful comments.
Finally, I acknowledge financial support
by Ministerio de Educaci\'on y Ciencia of
Spain under a contract Ram\'on y Cajal, and by CICYT of Spain under projects
FPA2002-01161 and FPA2005-01963.

\end{document}